\documentclass[]{spie}

\usepackage[]{graphicx}
\usepackage{amsmath}
\usepackage{amssymb}
\usepackage[]{aas_macros}
\usepackage{wrapfig}
\usepackage{bm}
\usepackage{ulem}
\usepackage{cancel}

\def\ra{\mathrm{a}}

\def\rc{\mathrm{c}}

\newcommand{\rd}{\mathrm{d}}

\def\bE{{\bf E}}

\def\rG{\mathrm{G}}

\def\bk{\boldsymbol{k}}
\def\rk{\mathrm{k}}

\def\rm{\mathrm{m}}
\def\bm{{\bf m}}

\def\bp{\boldsymbol{ p}}
\def\rp{\mathrm{p}}

\def\bQ{{\bf Q}}
\def\br{\boldsymbol{r}}
\def\rr{\mathrm{r}}

\def\rT{\mathrm{T}}
\def\ru{\mathrm{u}}

\def\balpha{\boldsymbol{\alpha}}

\def\brho{\boldsymbol{\rho}}
\def\btheta{\boldsymbol{\theta}}

\title{A Statistical Framework for the Utilization of Simultaneous Pupil Plane and Focal Plane Telemetry for Exoplanet Imaging, Part I: Accounting for Aberrations in Multiple Planes}

\author{Richard A. Frazin
\skiplinehalf
{\small Dept. of Climate and Space Sciences and Engineering, University of Michigan, Ann Arbor, MI 48109} }
\authorinfo{E-mail: rfrazin {\it \_at\_} umich.edu}

\begin{document}
\maketitle

\begin{abstract}

A new generation of telescopes with mirror diameters of 20 m or more, called extremely large telescopes (ELTs) has the potential to provide unprecedented imaging and spectroscopy of exo-planetary systems, if the difficulties in achieving the extremely high dynamic range required to differentiate the planetary signal from the star can be overcome to a sufficient degree.
Fully utilizing the potential of ELTs for exoplanet imaging will likely require simultaneous and self-consistent determination of both the planetary image and the unknown aberrations in multiple planes of the optical system, using statistical inference based on the wavefront sensor and science camera data streams.  
This approach promises to overcome the most important systematic errors inherent in the various schemes based on differential imaging, such as ADI and SDI.
This paper is the first in a series on this subject, in which a formalism is established for the exoplanet imaging problem, setting the stage for the statistical inference methods to follow in the future.
Every effort has been made to be rigorous and complete, so that validity of approximations to be made later can be assessed.
Here, the polarimetric image is expressed in terms of aberrations in the various planes of a polarizing telescope with an adaptive optics system.
Further, it is shown that current methods that utilize focal plane sensing to correct the speckle field, e.g., electric field conjugation, rely on the tacit assumption that aberrations on multiple optical surfaces can be represented as aberration on a single optical surface, ultimately limiting their potential effectiveness for ground-based astronomy.

\end{abstract}


\section{Introduction and Motivation}

The coming decades will see the construction and operation of a new class of ground-based telescopes with mirror diameters of 20 m or more.  Such instruments are called Extremely Large Telescopes (ELTs), and one of the top science priorities of the ELTs is direct imaging and spectroscopy of exoplanetary systems.
ELTs should allow direct imaging of Earth-like planets \cite{Guyon_habitable_ELT11}, but achieving this goal will require unprecedented precision in characterization of the optical system, as it is necessary to separate the starlight from the vastly fainter planetary emission.
The most promising method for achieving high-contrast imaging and spectroscopy from the ground combines high-order adaptive optics (AO) with a stellar coronagraph.
Typically, the AO system operates at visible wavelengths, while the science camera in the coronagraph captures images in the near-infrared.  Observing in the near-IR improves the needed contrast ratios and is required for spectroscopy of key molecules such as water and methane. 
(For a review of high-contrast direct imaging, the reader may consult [\citenum{Traub_Review}].)
The major limitation in high-contrast imaging is the appearance of so-called "quasi-static" speckles (QSS) in the image that can be brighter than the planetary emission.\cite{Boccaletti04, Martinez13}
These speckles are caused by constructive interference of the starlight, which has been distorted by aberrations in the optical system. 
QSS change on wide range of timescales due to a variety of mechanical stresses on the telescope caused by, for example, winds and temperature gradients, and they may last for days.

The current paradigm for treating the QSS is background subtraction based on differential imaging methods, the most important of which are angular differential imaging (ADI) and spectral differential imaging (SDI).
ADI and SDI have number of biases due to complicated self-subtraction issues.\cite{Marois_SOSIE,Rameau_ADI_SDI_limits15}
The self-subtraction problems make these methods unsuitable for true imaging, and they create many challenges for characterizing point sources (planets), resulting in a large and ever-growing literature.
As the basis of the SDI method is scaling of telescope point spread function (PSF) with wavelength, self-subtraction makes the method unsuited to imaging objects that are extended in the radial direction, and the method is sensitive to assumptions about the spectrum of the object. 
SDI can work well at large separations between the planet and star, but performs poorly at small separations since the wavelength-stretching effect is proportional to the distance from the center of the image.
This is unfortunate because the expected success in planet discovery is heavily dependent on being able to make observations at small star-planet separations.\cite{Stark_ExoEarthYield14,Brown_PlanetSearch15}
ADI relies on the diurnal field rotation to separate the PSF from the planetary emission, so that any circularly symmetric component in the image will be indistinguishable from the PSF and be subject to self-subtraction.
Combining SDI and ADI does not eliminate the self-subtraction problem, and resulting response pattern is rather complex.\cite{Rameau_ADI_SDI_limits15}   
Futher, SDI and ADI are plagued by a problem that the community has come to appreciate only very recently, which is that the uncertainties have been greatly under-estimated due to the fact that small sample statistics have not been taken into account.
Briefly, the background-subtracted image has residual speckles, whose amplitudes decrease with distance from the star.  
In order to determine an uncertainty level as a function of radius, the amplitude of the speckles as a function of radial distance is estimated from the background-subtracted image.
However, speckles have a size of $\sim \lambda/d$ (where $\lambda$\ is the wavelength and $d$\ is the telescope diameter), so, close to the host star there are very few speckles in a given small annulus.
These small sample considerations lead to a statistical penalty that exponentially increases toward small separations.\cite{Mawet_SmallNumStatsSpeckle14}
 
As the technology advances and higher levels of contrast are achieved, new problems associated with background subtraction via differential imaging are likely to surface.
For example, if the star exhibits a small amount of linear polarization, the point spread matrix (PSM), which is the Stokes-vector generalization of the PSF  [\citenum{Breckinridge15}], will depend on the diurnal rotation angle, further undermining the performance of ADI.  
Similar considerations apply to the planetary emission itself (which may be strongly linearly polarized due to Rayleigh scattering) if the PSM is not diagonally dominant.  
Furthermore, the PSFs of coronagraphic systems are sensitive to the pointing error, so that if the star is moving slightly relative to the pointing center during the diurnal rotation, the PSF will be changing, again, undermining the performance of ADI.

It is with this state of affairs in mind that several authors began to explore the utility of post-processing the pupil-plane telemetry, provided by a wavefront sensor (WFS), in conjunction with simultaneous image plane telemetry from a science camera, to treat the effects of the QSS.\cite{Frazin13,Codona13,Frazin14}
The historical context of this approach is reviewed in [\citenum{Frazin13}].
As the WFS must run at a frequency of approximately 1 kHz for the AO system to keep track of the atmospheric fluctuations, these methodologies also require $\sim1$ kHz exposure cadence in the science camera, a prospect made practical by a new generation of ultra-low noise IR cameras capable of kHz readouts, such as the SWIR single photon detector, SAPHIRA  eAPD and the MKIDS.\cite{SWIR_detector14,SELEX_APD12,Saphira_eAPD14,Mazin_MKIDS14}

\begin{figure}[t]
\includegraphics[width=.45\linewidth,clip=]{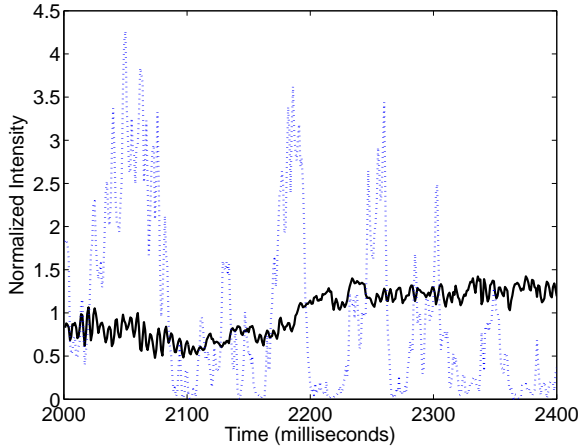}
\caption{\small Modulation of intensities at a single pixel (corresponding to the planet location) of the science camera in a stellar coronagraph simulation.  The black solid line shows the time-series of the temporal variation of the planetary intensity.  The dotted blue line shows the corresponding stellar speckle intensity at the same pixel.  Both the planetary and stellar intensity are normalized to have a mean of unity in this figure.  The stellar intensity is enhanced by a sinusoidal static aberration at the spatial frequency corresponding to the planet's location.  From [\citenum{Frazin13}], where this effect is demonstrated mathematically.}
\label{fig_modulation}
\vspace{-4mm}
\end{figure}

The combination of millisecond pupil plane and focal plane telemetry may well lead to orders of magnitude in contrast improvement over ADI and SDI, as it leverages a vastly larger and richer data set than standard exposure times, which average over the atmospheric turbulence.
It is important to emphasize that the millisecond imaging techniques can be generalized to take advantage of essentially all constraints on problem proposed to date.   
These constraints include those imposed by diurnal rotation (used by ADI), multi-wavelength observations (used by SDI), as well polarization (used in polarization differential imaging [\citenum{Hinkley_PDI09}]).
Thus, the potential losses of information are rather limited.
When one takes images that average over the atmospheric turbulence, the stellar speckles look much like planets.
But, with an AO-corrected signal,  planets and stellar speckles behave much differently at millisecond timescales.
This is illustrated in Fig.~\ref{fig_modulation}, which shows a stellar coronagraph simulation result from [\citenum{Frazin13}].
Fig.~\ref{fig_modulation} is a plot of two time series of the intensity calculated at a single pixel, corresponding to the location of a simulated planet, of the science camera.  
The dotted blue curve illustrates how the atmospheric turbulence (with AO correction) modulates the stellar speckle at a cadence of 1 millisecond.  The solid black curve shows the much weaker modulation of the planetary light in the same pixel.  
These two time-series, the planetary intensity and the speckle intensity, are quite different in character, with the speckle having an approximately exponential probability density function (PDF), while the PDF of the planetary intensity is localized around its non-zero mean.\cite{Gladysz10,Frazin13}
This can be understood as follows: Much as that of the star, the planet's wavefront is stabilized by the AO system, as the flat part of the planet's wavefront is responsible for its intensity at this position in the image plane.  
However, the stellar speckle intensity at that location is entirely due to the random, non-flat part of the star's wavefront (the coronagraph removes the flat part) and, hence, it is much more volatile.  
In an actual measurement, the signal seen would be a weighted sum of these two curves, with the star having much more weight, necessitating the collection of many milliseconds of data and the utilization of statistical inference methods to separate them.
Taking advantage of the fleeting moments in which the starlight is reduced at the planet's location,  
\begin{figure}[t]
\includegraphics[width=.43\linewidth,clip=]{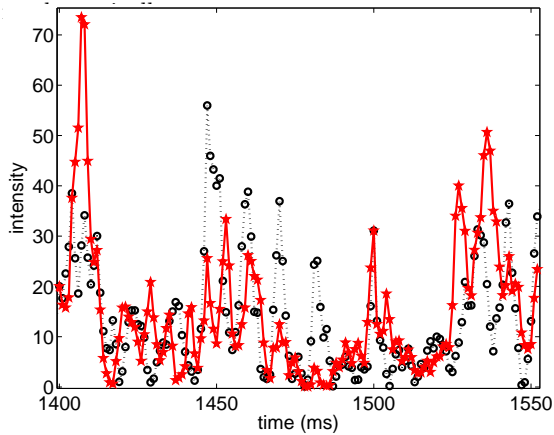} 
\caption{\small Stellar coronagraph simulation of the effect of a vibration on a speckle in the science camera.  This speckle is caused by a vibrational NCPA of the form $\cos(k \cdot r - \omega t)$, and the modulation is caused by the interaction of the NCPA with the random AO residual. The black dotted curve corresponds to the intensity of a speckle for $\omega/2\pi =100$ Hz and the red solid curve corresponds to $\omega/2\pi = 10$Hz.  The ms data are clearly sensitive to the frequency of the vibration.  Both curves are calculated with the same sequence of residual phase values.  From [\citenum{Frazin14}]. }
\label{Vibration}
\vspace{-3mm}
\end{figure}
\noindent was first considered by Labeyrie  [\citenum{Labeyrie_DarkSpeckle95}].

The richness of the millisecond telemetry is further illustrated by Fig.~\ref{Vibration}, taken from [\citenum{Frazin14}], which shows its sensitivity to high-frequency vibrations.  (Vibrations pose a particularly challenging problem to high-contrast astronomy.) 
Fig.~\ref{Vibration} shows the time-dependence of the intensity of a vibrational speckle caused by a pupil-plane aberration of the form $\phi(r) = \alpha \cos(\bk \cdot \br + \omega t)$, where $\omega$\ is the vibration frequency, $t$ is the time, $\bk$\ is the spatial frequency of the aberration and $\br$\ is the 2D spatial coordinate in the pupil plane.
The black and red curves correspond to two different frequencies (10 and 100 Hz), and the fact that they do not coincide shows that the millisecond data are sensitive to the frequency.   
Obviously, conventional observations that use exposures orders of magnitude longer would see the same speckle, quite independently of frequency.
Similarly, [\citenum{Frazin14}] showed that when a pupil plane aberration is given by $\phi(r) = \alpha \cos(k \cdot r + \vartheta)$, the millisecond data are sensitive to whether $\alpha$\ is real or imaginary as well as the phase angle $\vartheta$, while conventional imaging cannot distinguish these effects.

In  [\citenum{Frazin13,Frazin14}], the author's simulations use the millisecond telemetry from the focal and image planes to inform a large system of equations that provide a simultaneous statistical inference of both the aberrations in the optical system and the planetary image, avoiding background subtraction altogether.
In contrast, in the background estimation method of  [\citenum{Codona13}], the millisecond telemetry is used to solve for the electric field due to the aberrations, as will be discussed later in Sec.~\ref{Methods}\ref{BlackBox}.
It is important to note that the methods of  [\citenum{Frazin13}] and  [\citenum{Codona13}] are fully compatible with utilizing the same constraints that ADI and SDI, as well speckle nulling and electric field conjugation methods  [\citenum{Traub_Nulling06,Pueyo_EFC09}], which themselves will be discussed in Sec.~\ref{Methods}.

The objective of this series of papers is to provide a rigorous statistical inference framework that is suitable for the high-precision requirements of exoplanet imaging with ELTs that are capable millisecond telemetry in both the pupil and image planes. 
The developments will account for polarization effects and detailed propagation within the optical system, noise and bandwidth effects in the wavefront sensor, as well as noise in the polarimetric image measured by the science camera.

The purpose of this paper, the first in the series, is to provide a formal description of polarimetric formation of an astronomical image in the presence of atmospheric turbulence, allowing for vector propagation of the wavefront through an adaptive optics telescope system with unknown aberrations at various surfaces in the optical system. 
The developments here also provide a new perspective on current methods that utilize focal plane sensing to treat the unwanted speckle field.
  Subsequent papers in this series will build upon this foundation to treat statistical inference of the aberrations and planetary image while treating practical limitations of the hardware.

\section{Elementary Concepts}\label{ElemConc}

The portion of the astronomical community with an interest in precise optical measurements, including those requiring ultra-high contrast, is increasingly becoming aware of the importance of polarization changes imparted to the light by the various elements of the telescope's optical system.
These (often undesired) changes are generally called ''polarization aberration," and are most commonly caused by reflection off mirrors at non-normal incidence.
The effects of polarization aberration are clearly illustrated by the point-spread matrix in [\citenum{Breckinridge15}], which operates on the Stokes vector of the incoming light.
This section introduces a straightforward formalism to treat polarimetric imaging system operating on a wave that is modulated by atmospheric turbulence.  
In addition, notations and conventions to be used throughout this series of papers are introduced here.

\subsection{timescales and Coherence Functions}\label{TimeScales}

The developments presented below will involve three timescales.
The smallest timescale is the coherence time $\tau_\mathrm{c}$\ of thermal light that has gone through a bandpass filter that removes all of the light except within a band of width $\Delta \nu$\ centered on frequency $\nu$, i.e., $\tau_\mathrm{c} = 1/\Delta \nu$.\cite{StatisticalOptics}  
For example, setting the wavelength $\lambda = 2 \mu$m and a narrow bandpass of just 1\%, $\tau_\mathrm{c} \approx 6.7\times10^{-13}$\ s.
This timescale is relevant in the definition of the coherence functions, which require averaging over many periods of $\tau_\mathrm{c}$\ so that these functions obtain their mean values or very nearly so.  
The middle timescale, here referred to as the "Greenwood time," $ \tau_\mathrm{G} = 1/\nu_\mathrm{G}$, where $\nu_\mathrm{G}$\ is the Greenwood frequency.
The Greenwood time, on the order of $10^{-3}$ s, is the time period in which most of the atmospheric phase distortions change significantly and defines the bandwidth requirement for an effective AO system.\cite{Greenwood77}
The largest timescale, $\tau_\mathrm{d}$ is that for which the telescope optical systems exhibit dynamical aberration (due to time-variable mechanical stresses) that change the detailed structure of the image in the science camera.  
Precise values of these three timescales are of little importance in the developments below so long as they satisfy $\tau_\mathrm{c}<< \tau_\mathrm{G} << \tau_\mathrm{d}$, although the possibility of using millisecond observations to detect rapid mechanical vibrations, perhaps even with periods approaching $10^{-3}$\ s is an intriguing one that is worth further investigation.\cite{Frazin14}

Any spatially variant time-harmonic field can be represented in terms of an elliptically polarized wave with the electric field vector confined to some plane. 
However, in a field with arbitrary spatial variation, the normal vector specifying this plane will be a function of position, as will the polarization parameters.\cite{Born&Wolf}
In Sec.~\ref{PropModels}\ref{atmosphere} it will be argued that atmospheric phase modulation will cause the normal vector describing the plane of polarization to wobble slightly on time-scale of $\tau_\rG$.
Rigorously accounting for this effect is beyond the scope of this paper and is left to future work.
Thus, it will be assumed that the electric field component of the light that the telescope is collecting can be represented as a quasi-monochromatic signal admitting a Jones vector representation in a known plane.\cite{Born&Wolf,Collett93}
Consider the following representation of the analytic signal  [\citenum{Born&Wolf,StatisticalOptics}] corresponding to the amplitudes of the electric field fluctuations in orthogonal directions of a plane-wave propagating in the $+z$\ direction :
\begin{align}
\bE(t)  = &
\left[ \begin{array}{l}
E_x(t) \exp j [ \Delta'_x(t) + \varsigma_x(t)]  \\ 
E_y(t) \exp j [ \Delta'_y(t) + \varsigma_y(t)]
\end{array}
 \right]  \label{E_complicated} \\
 \equiv &
 \left[ \begin{array}{l}
u_x(t)   \\ 
u_y(t) 
\end{array}
 \right] 
\equiv u(t)   \, , 
\end{align}
which defines the Jones vector $u$\ to be used throughout this presentation, and where a factor of $\exp j[kz -2 \pi \nu t]$, with $k = 2 \pi / \lambda $, has been suppressed.
Constraints on time-dependent functions in Eq.~(\ref{E_complicated}) are described below.
While $E_x(t)$\ and $E_y(t)$ are in general complex-valued, $ \Delta'_x(t), \; \Delta'_y(t), \; \varsigma_x(t)$\ and $\varsigma_y(t)$\ can be taken to be real.
The random processes  $\varsigma_x(t)$\ and $\varsigma_y(t)$\ are zero-mean so that $\langle \varsigma_x(t) \rangle_{\tau_\rc} = \langle \varsigma_y(t) \rangle_{\tau_\rc} = 0$, where the brackets $\langle \, \rangle_{\tau_\mathrm{c}}$\ indicate a time average in which, for some function $f(t)$:
\begin{equation}
\langle f(t) \rangle_{\tau_c} \equiv 
\frac{1}{ \Delta t} \int_{t-\Delta t}^{t} \rd t' \, f(t') \, ,
\label{short_time_average}
\end{equation}
where $\tau_\mathrm{c} << \Delta t << \tau_\mathrm{G}$.  
Thus, $\langle f(t) \rangle_{\tau_c}$\ is a low-pass version of $f(t)$.   

The random processes $\varsigma_x(t)$\ and $\varsigma_y(t)$\ control the correlation between the $x-$ and $y-$\ components, and they vary on timescales on the order of $\tau_\rc$\ or longer, but much less than $\tau_\mathrm{G}$.
The functions $E_x(t)$, $E_y(t)$, $\Delta'_x(t)$\ and $\Delta'_y(t)$ are allowed to fluctuate on timescales on order of $\tau_\mathrm{G}$, but are quite constant over timescales similar to $\tau_\mathrm{c}$, so that, to a very high degree of accuracy, $E_x(t) = \langle E_x(t) \rangle_{\tau_c} $,  $E_y(t) = \langle E_y(t) \rangle_{\tau_c} $, $\Delta'_x(t) = \langle \Delta'_x(t) \rangle_{\tau_c} $
 and $\Delta'_y(t) = \langle \Delta'_y(t) \rangle_{\tau_c} $.
In this way, the high-frequency character of the optical field has been placed into the $\exp(j 2 \pi \nu t)$, $\exp(j \varsigma_x(t))$ and $\exp(j \varsigma_y(t))$\ factors.
$E_x(t)$\ and $E_y(t)$\ are required to be in phase, so that 
\begin{equation}
E_x(t) = |E_x(t)| e^{j \phi(t)} \: \: \: \mathrm{and} \: \: 
E_y(t) = |E_y(t)| e^{j \phi(t)} \, ,
\label{in-phase}
\end{equation}
which have the same phase factor $\phi(t)$.
It will be useful to factor $u(t)$\ into two parts, a scalar component representing the commonalities between $u_x$\ and $u_y$\ and a polarization state representing their differences:
\begin{equation}
u(t) = \underbrace{\sqrt{ I(t) \; }\exp[ j \phi(t)]}_{\mathrm{scalar \: component}} 
 \underbrace {\left[ \begin{array}{l}
e_x(t) \exp j [ \Delta_x(t) + \varsigma_x(t)]  \\ 
e_y(t) \exp j [ \Delta_y(t) + \varsigma_y(t)]
\end{array}
 \right] }_{\mathrm{polarization \: state}} \, , 
\label{polariz_factor}
\end{equation}
where $I(t) = \sqrt{ E_x(t) E^*_x(t) + E_y(t)E^*_y(t) \, }$\ is the intensity, $0 \leq e_x(t), e_y(t) \leq 1$, $ e^2_x(t) + e^2_y(t) = 1$, $\phi(t) = (\Delta'_x(t) + \Delta'_y(t))/2 $, so that $\Delta_x(t) + \Delta_y(t) = 0$, and $^*$\ indicates complex conjugation.
The form of Eq.~(\ref{polariz_factor}) is convenient for easy interpretation.  
The scalar component controls the complex amplitude of the electric field vector,
and in the polarization state, $e_x(t)$\ and $e_y(t)$\ set the relative amplitudes of the $x-$ and $y-$\ components.
$\Delta_x(t) - \Delta_y(t)$ determines linear vs. circular polarization.
The degree of polarization is set by $ e_x(t), \, e_y(t) , \, \varsigma_x(t)$\ and $\varsigma_y(t)$.   

For an astronomical signal just above the Earth's atmosphere, $I, \, e_x, \, e_y, \, \phi, \, \Delta_x$\ and $\Delta_y$\ are essentially constant in time, so the factorization in Eq.~(\ref{polariz_factor}) becomes:
\begin{equation} \begin{split}
u(t) & = \sqrt{I \;} \exp[ j \phi] 
 \left[ \begin{array}{l}
e_x \exp j [ \Delta_x + \varsigma_x(t)]  \\ 
e_y \exp j [ \Delta_y + \varsigma_y(t)]
\end{array}
 \right]  \\
 &  \equiv \sqrt{I \;} \exp[ j \phi] \breve{u}(t) 
 \, , 
\label{polariz_factor2}
\end{split} \end{equation}
thus defining the polarization state vector $\breve{u}(t)$.
The time-dependent effects of the atmosphere and the telescope system will be assumed to be representable as linear operators, which take the form of a Jones matrix, as described below in Sec.~\ref{ElemConc}.\ref{ImageFormation}.
The fluctuations related to atmospheric and telescopic dynamics on the timescales of $\tau_\mathrm{G}$\ and $\tau_\mathrm{d}$, respectively, do not cause depolarization, but they can cause other changes in the polarization.
When an astronomical signal of the form in Eq.~(\ref{polariz_factor2}) is multiplied by a time-dependent Jones matrix that varies on timescales of $\tau_\mathrm{G}$\ or longer, the resulting field can be described by the more general form in Eq.~(\ref{polariz_factor}).

Describing the intensities measured by polarization sensitive instruments requires the 2\underline{nd} order statistics of $u$, which are contained in the $4 \times 1$ coherency vector (usually called the coherency matrix when arranged in the form of a $2\times 2 $ matrix),\cite{Born&Wolf,StatisticalOptics} and is given by the time-averaged Kronecker product of $u$ and $u^*$, denoted as $u \otimes u^*$  [\citenum{McGuire90}]:
\begin{align}
J(t)  &= \langle u(t) \otimes u^*(t)  \rangle_{\tau_\rc}  \nonumber \\
& =  I(t) \exp \left[ - 2 \Im \big( \phi(t)  \big) \right] 
  \left[  \begin{array}{l}
e^2_x(t) \\
e_x(t)e_y(t)\exp j [ \Delta_x(t) - \Delta_y(t)  ] \langle \exp j [ \varsigma_x(t) - \varsigma_y(t)  ]  \rangle_{\tau_\mathrm{c}} \\
e_y(t)e_x(t)\exp j [ \Delta_y(t) - \Delta_x(t)  ]  \langle \exp j [ \varsigma_y(t) - \varsigma_x(t)  ] \rangle_{\tau_\mathrm{c}} \\
e^2_y(t)   \\ 
\end{array} \right]  \, .
\label{coherency_def}
\end{align}
where the $\Im$\ symbol indicates the imaginary part.
The second equality in Eq.~(\ref{coherency_def}) makes explicit use timescale separations explained above.
In order that the linear vs. circular polarization nature of the light is set only by $\Delta_x(t) - \Delta_y(t)$, it is clear that the condition $ \langle \varsigma_x(t) - \varsigma_y(t) \rangle_{\tau_\rc} = 0$\ must be satisfied, as any non-zero mean would modify the relative phase of the $x-$\ and $y-$components of the field.

Most intensity measurements are more readily interpreted in terms of the Stokes vector, which is easily obtained from $J$:
\begin{equation}
J^\mathrm{S}(t) = \bQ J(t)  \; , \:  \mathrm{where} \: \: \: 
\bQ = \left[ \begin{array}{cccc}
1 & 0 & 0 & 1\\
1 & 0 & 0 & -1\\
0 & 1 & 1 & 0\\
0 & -j& j & 0\\
\end{array} \right] \, ,
\end{equation}
where the 0\underline{th} component of $J^\mathrm{S}$ is the total intensity, the 1\underline{st} corresponds to vertical{/}horizontal polarization, the 2\underline{nd} to diagonal polarization and the 3\underline{rd} to circular.\cite{Born&Wolf,Collett93}
Note that $\bQ$\ is invertible and has 4 eigenvalues that all have an absolute value of $\sqrt{2}$. 
Thus, the coherency and Stokes vectors are equivalent representations, and it follows that a measurement of the full Stokes vector fully specifies the beam, allowing determination of $I (t) \exp \big[ -2 \Im \big( \phi(t) \big) \big] $, $e_x(t)$, $e_y(t)$, $\Delta_x(t) - \Delta_y(t)$, and $ \langle \exp j [ \varsigma_x(t) - \varsigma_y(t)  ]  \rangle_{\tau_\mathrm{c}}$.

When considering wave propagation, it is necessary to allow the field to have spatial dependence in the transverse plane.   
Let $\br_1$\ and $\br_2$\ be the transverse [i.e., $(x,y)$] coordinates of two points in some plane, and let $t_1$ and $t_2$ be two times.
The $4 \times 1$\ cross-density vector, $\Gamma(\br_1,\br_2,t_1,t_2) $, is a generalization of the coherency ([\citenum{Wolf03}] calls this the "cross-spectral density").
In the most general case, the $4 \times 1$\ cross-density vector is defined as $\Gamma(\br_1,\br_2,t_1,t_2) \equiv  \langle u(\br_1,t_1) \otimes u^*(\br_2,t_2) \rangle_{\tau_\mathrm{c}} $, so that $\Gamma(\br,\br,t,t) = J(\br,t)$.
For the purposes of image formation it is sufficient to only allow time differences $\tau \equiv t_2 - t_1$\ such that $| \tau | < \tau_\rc$.
As the $\varsigma_x$\ and $\varsigma_y$\ stochastic processes were assumed to be stationary, the rapidly fluctuating part of the cross-density (i.e., at timescales $<< \tau_\rG$) can only be a function of the time difference $\tau$.\cite{StatisticalOptics}
Further, as the fields are taken to be quasi-monochromatic, $\Gamma(\br_1,\br_2,t, t + \tau) = \langle u(\br_1,t) \otimes u^*(\br_2,t + \tau) \rangle_{\tau_\mathrm{c}} = \langle u(\br_1,t) \otimes u^*(\br_2,t ) \rangle_{\tau_\mathrm{c}} \exp( - j 2 \pi \nu \tau)$.
The mutual coherency $\gamma$\ is defined as the cross-density evaluated at 0 time difference, so,  $\gamma(\br_1,\br_2,t) \equiv  \Gamma(\br_1,\br_2,t, t) = \langle u(\br_1,t) \otimes u^*(\br_2,t ) \rangle_{\tau_\mathrm{c}}$.
As will be seen in Sec.~\ref{ElemConc}\ref{ImageFormation}, the mutual coherency plays an important role in polarimetric image formation.

Consider the field of an astronomical source, as represented in Eq.~(\ref{polariz_factor2}),
at the spatial location $\br$\ in some plane just above the Earth's atmosphere.
Clearly, $I$, $e_x$, $e_y$, $\Delta_x$ and $\Delta_y$\ must be independent of $\br$\ because the Stokes parameters that would be measured in that plane (at least not too far from Earth) are independent of $\br$.  
But, in Sec.~\ref{PropModels}\ref{ToEarth} it will be seen that the mutual coherency $\gamma(\br_1,\br_2)$\ carries an encoding of the image of the astronomical source.
Therefore, the image information must be contained the functions $\phi$, $\varsigma_x$\ and $\varsigma_y$, which must be functions of the spatial coordinate $\br$.
To see this, consider the definition of the mutual coherency.
Similarly to Eq.~(\ref{coherency_def}), the mutual coherency is given by:
 \begin{equation}
\begin{split}
\gamma(\br_1,\br_2)& = \langle u(\br_1,t) \otimes u^*(\br_2,t)  \rangle_{\tau_\rc}  \\
 & = I \exp j \big[ \phi(\br_1) - \phi(\br_2) ] \; \times \\
 & \left[ \begin{array}{l}
e_x^2 \langle \exp j [ \varsigma_x(\br_1,t) - \varsigma_x(\br_2,t)  ] \rangle_{\tau_\mathrm{c}}  \\
e_x e_y \exp j [ \Delta_x - \Delta_y  ] \langle \exp j [ \varsigma_x(\br_1,t) - \varsigma_y(\br_2,t)  ]  \rangle_{\tau_\mathrm{c}} \\
e_y e_x \exp j [ \Delta_y - \Delta_x  ]  \langle \exp j [ \varsigma_y(\br_1,t) - \varsigma_x(\br_2,t)  ] \rangle_{\tau_\mathrm{c}} \\
e_y^2 \langle \exp j [ \varsigma_y(\br_1,t) - \varsigma_y(\br_2,t)  ] \rangle_{\tau_\mathrm{c}}   \\ 
\end{array} \right]  \, ,
\end{split}
\label{mutual_coherency_def}
\end{equation}
where the time argument of $\gamma$\ has been dropped because it does not depend on time for astronomical signals above the atmosphere.
As $\phi(\br)$\ is strictly real above the atmosphere, it has no influence on the amplitude of $\gamma(\br_1,\br_2)$.
Thus, the amplitude of $\gamma(\br_1,\br_2)$\ is determined by the function $ \langle \exp j [ \varsigma_x(\br_1,t) - \varsigma_y(\br_2,t)  ] \rangle_{\tau_\mathrm{c}} $\ and its three cousins in Eq.~(\ref{mutual_coherency_def}).

According to the Van Cittert-Zernike theorem for scalar fields , the astronomical image can be determined by taking the Fourier transform of the mutual coherence function and [\citenum{StatisticalOptics}].
Since vector analog of the mutual coherence function is the mutual coherency in Eq.~(\ref{mutual_coherency_def}), the polarimetric image of the astronomical source is encoded in $\phi(\br_1) - \phi(\br_2)$ (which is analogous to the phase of the mutual coherence function), and the functions $\langle \exp j [ \varsigma_x(\br_1,t) - \varsigma_x(\br_2,t)  ] \rangle_{\tau_\mathrm{c}}$, $\langle \exp j [ \varsigma_x(\br_1,t) - \varsigma_y(\br_2,t)  ] \rangle_{\tau_\mathrm{c}}$ and $\langle \exp j [ \varsigma_y(\br_1,t) - \varsigma_y(\br_2,t)  ] \rangle_{\tau_\mathrm{c}}$ (which are analogous to the magnitude of the mutual coherence function, also known as the "visibility" in long-baseline stellar interferometry).
This will be discussed further in Sec.~\ref{PropModels}\ref{ToEarth}.
As discussed after Eq.~(\ref{coherency_def}), $\varsigma_x(\br,t)$\ and $\varsigma_y(\br,t)$ must satisfy the condition $\langle \varsigma_x(\br,t) - \varsigma_y(\br,t) \rangle_{\tau_\rc} = 0$, but the value of $\langle \varsigma_x(\br_1,t) - \varsigma_y(\br_2,t) \rangle_{\tau_\rc} $\ has no such restriction on its behavior (and similarly for its $x,x$ and $y,y$ cousins).  
To understand this, consider an extended astronomical source with purely linear $45^\circ$\ polarization, so ($e_x=e_y=1/\sqrt{2}$), i.e., the coherency of the light emerging from the source is given by $S(\balpha) = s(\balpha) S$, in which $\balpha$\ corresponds to the position in the sky, $s(\balpha)$ is a real-valued scalar function, and the vector $S = \bQ^{-1}[1,0,1,0]^\rT$ (where $^\rT$\ indicates transpose).
In this case, the source has a spatially constant polarization, so the mutual coherency seen at the Earth must exhibit scalar behavior, i.e., $\gamma(\br_1,\br_2) = g(\br_2 - \br_1)J$, where $J$\ is a constant coherency vector and $g(\br)$\ is some scalar function.
The scalar behavior of $\gamma$\ requires $\langle \varsigma_x(\br_1,t) - \varsigma_x(\br_2,t)  \rangle_{\tau_\mathrm{c}} = \langle \varsigma_x(\br_1,t) - \varsigma_y(\br_2,t)  \rangle_{\tau_\mathrm{c}} = \langle \varsigma_y(\br_1,t) - \varsigma_y(\br_2,t)  \rangle_{\tau_\mathrm{c}} = 0$. 
Further, one has $\langle \exp j [ \varsigma_x(\br_1,t) - \varsigma_x(\br_2,t)  ] \rangle_{\tau_\mathrm{c}} =\langle \exp j [ \varsigma_x(\br_1,t) - \varsigma_y(\br_2,t)  ] \rangle_{\tau_\mathrm{c}} = \langle \exp j [ \varsigma_y(\br_1,t) - \varsigma_x(\br_2,t)  ] \rangle_{\tau_\mathrm{c}} = \langle \exp j [ \varsigma_y(\br_1,t) - \varsigma_y(\br_2,t)  ] \rangle_{\tau_\mathrm{c}}$.
However, for realistic sources in which the polarization state is not spatially constant, these conditions do not hold, allowing the "polarization ellipse," if it were ever defined for the mutual coherency, to change as a function of $\br_1 - \br_2$.

\subsection{Polarimetric Image Formation and Polarization Aberration}\label{ImageFormation}

One may model the effect of a perfectly flat optical surface, such as an ideal flat mirror, on a plane wave with a $2\times2$\ Jones matrix $\Upsilon$ operating on the Jones vector $u$, so that the outgoing plane wave is given by $ u' = \Upsilon u$.\cite{Collett93}
The coherency vector of the outgoing wave is given by ${J'} = (\Upsilon u)  \otimes (\Upsilon^* u^*)  = (\Upsilon \otimes \Upsilon^*)(u \otimes u^*) = (\Upsilon \otimes \Upsilon^*) J$, which makes use of the common identity for Kronecker products of matrices: $AC \otimes BD = (A\otimes B)(C \otimes D)$.\cite{Horn&Johnson2}
As explained in the fine paper by McGuire \& Chipman [\citenum{McGuire90}], so long as the paraxial approximation applies, this result can be extended to any optical system that is non-scattering and does not depolarize the light.\cite{Breckinridge15}
Let the field in entrance pupil plane, denoted with index $0$, of some such optical system be represented by the Jones vector $u_0(\br_0,t)$, where $\br_0$\ is the 2D coordinate vector in the plane, and let the outgoing field in the exit plane (with index number $1$) be represented by $u_1(\br_1,t)$.
For example, plane 0 could correspond to the entrance pupil of a telescope and the plane 1 could correspond to the surface of a detector where an image is formed.
The optical system is modeled in terms of a Jones propagation matrix $\Upsilon(\br_1,\br_0)$:
\begin{equation}
u_1(\br_1,t) = \int \rd \br_0 \, \Upsilon(\br_1,\br_0) u_{0}(\br_0,t) \; ,
\label{JonesPupil}
\end{equation}
where $\Upsilon(\br_1,\br_0)$\ is 0 when $\br_0$\ is not within the spatial limits of the entrance pupil.
At first glance, Eq.~(\ref{JonesPupil}) implies instantaneous action and violates the laws of relativity, however, the time delays in the optical system are accounted for by phase shifts, such as a quadratic phase term describing a lens or focusing mirror.\cite{IntroFourierOptics}
As per the conventions {\it mise en place} in the Appendix, the $t$\ in Eq.~(\ref{JonesPupil}) only refers to fluctuations on timescales far longer than $\tau_\rc$\ and the action can be considered to be instantaneous at such timescales.  
The coherency vector of the outgoing light is given by:
\begin{equation}
J_1(\br_1,t)  =    u_1(\br_1,t) \otimes u^*_1(\br_1,t)
 = \int \rd \br_0 \int \rd \br_0'   
 \left[\Upsilon(\br_1,\br_0) \otimes \Upsilon^*(\br_1,\br_0') \right]  \gamma_{0}(\br_0,\br_0',t) \, ,
\label{JonesPupilKron}
\end{equation}
where $\gamma_0$ is the mutual coherency in plane $0$.
The $4 \times 4$\ matrix of functions $\Upsilon(\br_1,\br_0) \otimes \Upsilon^*(\br_1,\br_0')$\ is the propagation kernel for the mutual coherency.
Thus, the coherency vector output by the optical system can be expressed in terms of the Jones propagator of the optical system and the mutual coherency vector at the entrance pupil.\cite{Breckinridge15}

Below [in Eq.~(\ref{vCZ})], it will be seen that $\gamma $\ is essentially the spatial Fourier transform of the scene to be imaged.
Therefore, if $\Upsilon$\ in Eq.~(\ref{JonesPupilKron}) has Fourier transforming properties, then $J_1(\br,t)$\ will be a polarimetric image of the scene.
As is emphasized in [\citenum{Breckinridge15}], only an optical system in which $\Upsilon \otimes \Upsilon$\ is proportional to the identity matrix will treat all Stokes parameters of the input beam the same way, and any deviation from this ideal is called polarization aberration.  
Examples of polarization aberration in telescopes are discussed in [\citenum{Breckinridge15}], and [\citenum{McGuire90}] analyzes a circular retarding lens (with corn syrup as the key ingredient) for which left and right circularly polarized light have different focal lengths.

\section{Propagation Models}\label{PropModels}

Eq.~(\ref{JonesPupilKron}) relates the polarimetric image produced by an imaging system to the mutual coherency of the optical radiation incident upon its entrance pupil.
The objective of this section to follow the mutual coherency of the light starting from the astronomical object to the Earth, through the atmosphere and, finally, through the telescope while accounting for unknown aberrations and the vector nature of the field.
The derivation given in Sec.~\ref{PropModels}\ref{ToEarth} for the polarimetric version of the van Cittert-Zernike theorem is similar to the scalar version in [\citenum{StatisticalOptics}].
Henceforth, this paper will use the notation conventions given in the Appendix.

\subsection{To Earth}\label{ToEarth}

Consider a planetary system, as viewed from the direction of Earth, and assume that the telescope is pointed somewhere close to the center of the system, where the host star is located.  
The pointing direction of the telescope corresponds to the $-z$\ direction, so that light from the planetary system collected by the telescope is traveling in more-or-less the $+z$\ direction.
Then, any point in the planetary system is given by the coordinates $(z_0 \balpha,-z_0)$, where $z_0$\ is the distance from the planetary system to Earth, and $\balpha$\ is a two-dimensional vector, with units of radians, corresponding to the direction cosines $-x/z_0$\ and $-y/z_0$.

At optical frequencies, outer space is effectively homogenous and isotropic, so there are no polarization effects and wave propagation is a scalar phenomenon.
Then, the Huygens-Fresnel principle can be used to propagate each component of $u$\ independently to a plane (normal direction $z$) just above the Earth's atmosphere.\cite{IntroFourierOptics}
Denote this plane with the index $-1$\ (below, the $0$ plane will correspond to the telescope entrance pupil, and the positive indices will correspond to subsequent planes within the optical system).
The optical radiation emerging from the planetary system can be considered to be emitted from the $z=-z_0$ plane, which, say, cuts through the star, and the field emerging from the plane is given by $u_\rp(z_0 \balpha)$.  
Then, the planetary system gives rise to the vector field $u_{\rp -1}(\br)$\ at location $\br$\ in the $-1$\ plane, just above Earth's atmosphere, which is given by  [\citenum{StatisticalOptics}]:
\begin{equation}
u_{\rp -1}\big( \br) = \frac{-k}{2 \pi} \int_\rp \rd (z_0^2  \balpha) \,
 u_\rp \left( z_0 \balpha  \right)  
 \frac{\exp[ -j k r(\br,\balpha)  ]}{r(\br,\balpha)}\cos\vartheta(\br,\balpha) \, ,
\label{u_p-1}
\end{equation}
where $u_\rp$\ is the field emerging from the planetary system, $ r^2(\br,\balpha) = z_0^2 + (\br - z_0 \balpha)^2 $, and $\cos\vartheta(\balpha,\br)$\ is the obliquity factor.

As the planetary system consists of many independently radiating sources, the emergent light is effectively spatially incoherent, so that the mutual coherency of the emergent radiation is given by $u_\rp(z_0 \balpha_1) \otimes u^*_\rp(z_0 \balpha_2) = (4 \pi / k^2) J_{\rp}(z_0 \balpha_1)  \delta(\balpha_2 - \balpha_1)$.\footnote{
The only truly spatially incoherent source corresponds to $\gamma = 0$, and [\citenum{StatisticalOptics}] advocates approximating the incoherent condition with the scale factor used here, corresponding to a field that is coherent over an area roughly the size of $\lambda^2$.
} 
Using the spatially incoherent source condition and the fact that $|z_0 | >> |\br| $, one finds:
\begin{eqnarray}
\gamma_{\rp -1}(\br_1,\br_2) & = & u_{\rp -1}(\br_1) \otimes u^*_{\rp -1}(\br_2)
\nonumber \\ & = &
\frac{ \exp [j p(\br_1,\br_2)]}{\pi }
\int_\rp   \rd  \balpha \,
J_\rp (z_0 \balpha) 
\exp \left[ j k \big(  \br_1 - \br_2 \big) \cdot \balpha  \right] ,
\label{vCZ-1}
\end{eqnarray}
where the quadratic phase factor is $p(\br_1,\br_2) = (k/2z_0)(\br_1 \cdot \br_1 - \br_2 \cdot \br_2)$.
For astronomical sources the $\exp [j p(\br_1,\br_2)]$\ factor can be dropped since the telescope diameter squared is much smaller than $2z_0/k$ (e.g., for a 100 m telescope operating at 1 $\mu$m and targeting {\it Alpha Centauri A}, this ratio is at most $\approx 10^{-6}$).

Note that in classical radiometry $J_\rp(z_0 \balpha)$\ has units of [energy/(area solid-angle time)] and is called the radiance, and $\rd \balpha$\ has units of [solid-angle].\cite{Born&Wolf}
In astronomy, the preference is for surface brightness, which also has units [energy/(area solid-angle time)], but is a function of the angular coordinate $\balpha$\ instead of the linear coordinate $(z_0 \balpha)$.\cite{AstrophysicalQuantities}
Define the planetary surface brightness coherency vector as $S(\balpha) \equiv J_\rp(z_0 \balpha)$ [note that this does not constitute a change of variables, so scaling by the Jacobian is not required], so that Eq.~(\ref{vCZ-1}) becomes:
\begin{equation}
\gamma_{\rp -1}(\br_1,\br_2) 
  =  \frac{1}{\pi}  \int_\rp   \rd \balpha \,
S(\balpha) 
\exp \left[ j k \big(  \br_1 - \br_2 \big) \cdot \balpha  \right] \, .
\label{vCZ}
\end{equation}
Eq.~(\ref{vCZ}) is the vector-field generalization of the well-known van Cittert-Zernike theorem for astronomical sources, and states that the mutual coherency vector of the light arriving at the Earth is proportional to the Fourier transform of the coherency vector of the light emerging from the planetary system, $S(\balpha)$.  
The objective of this paper is to develop a methodology for determining $S(\balpha)$\ in the high-contrast context.
Tervo et al. [\citenum{Tervo13}] derive the same result in terms of the Stokes parameters and discuss the history of previous similar efforts.

The coherency of the planetary light above the atmosphere is easily calculated from Eq.~(\ref{vCZ}) as $J_{\rp -1}(\br) = \gamma_{\rp -1}(\br,\br)$, resulting in:
\begin{equation}
J_\rp   = \frac{1}{\pi}  \int_\rp   \rd \balpha \, S(\balpha) \, .
\label{J_p-1}
\end{equation}  
Note that $J_\rp $\ has no dependence on the spatial coordinate $\br$.

In principle, Eq.~(\ref{vCZ}) describes the mutual coherency arising from the entire planetary system, central star included. 
However, in the high-contrast context the star is vastly brighter than the surrounding material and the propagation of its light must be treated with much more care than the planetary light.
Therefore, the starlight and planetary light are to be treated separately in these developments.
Assume that the star is unresolved, essentially acting as a point source, and is located at a small angle $\balpha_\star$\ from the telescope pointing direction, and let $J_\star $\ be its coherency vector, as would be measured above the atmosphere.
The star's mutual coherency can be calculated with the aid of Eq.~(\ref{vCZ}) by setting $S(\balpha) = \pi J_\star \delta(\balpha - \balpha_\star)$, resulting in:
\begin{equation}
\gamma_{\star -1}(\br_1,\br_2) = 
 J_\star \exp j \big[  k(\br_1-\br_2) \cdot \balpha_\star  \big] \, .
\label{VcZ_star}
\end{equation}
Since $\gamma_{\star -1}(\br_1,\br_2) = u_{\star -1}(\br_1) \otimes u^*_{\star -1}(\br_2) $, one can see that $u_{\star -1}$ can be factored as per Eq.~(\ref{polariz_factor2}):
\begin{equation}
u_{\star -1}(\br) = \sqrt{I_\star \,}
\exp (j k \balpha_\star \cdot \br   ) \breve{u}_\star \,
\label{u_star_above_atmos}
\end{equation}
where $I_\star$\ is star's intensity, $k \balpha_\star \cdot \br $, is the real-valued phase, and $\breve{u}_\star $\ is the polarization state vector.
With the variables so-defined, one has:
\begin{align}
J_\star &= \gamma_{\star -1}(\br,\br) =  u_{\star -1}(\br) \otimes u^*_{\star -1}(\br) \nonumber \\
& =  I_\star \breve{u}_\star \otimes \breve{u}^*_\star \, .
\label{J_star_above_atmos}
\end{align}

\subsection{Through the Atmosphere}\label{atmosphere} 

The journey from the $-1 $ plane (above the atmosphere) to the $0$ plane, corresponding to the entrance pupil of the telescope, requires the light to traverse the Earth's atmosphere.
While atmospheric polarization aberration has been discussed in the military context of propagation of beams of laser light ([\citenum{XiaolingJi09}] has a useful list of references) the literature on the subject in the context of astronomy appears to be rather limited.  
The most important effect of the atmosphere is that of a random phase screen, and the second most important effect is scintillation, which itself is caused by phase screening at larger heights.\cite{Roddier81}
Absorption effects by the atmosphere should only affect photometry and are generally well-understood.\cite{AstrophysicalQuantities}
The atmosphere generally not expected to produce polarization effects, however, several studies have found polarization related to presences of specific particles or molecules or molecules in the atmosphere.
For example,  [\citenum{Kemp_SolarPol08}] found small, wavelength-dependent linear polarization in high-precision solar observations at large zenith angles, and attributed it to double scattering off of various molecules.
Bailey et al.~ [\citenum{Bailey_SkyPol08}] report a slight linear polarization caused by Saharan dust high in the atmosphere over the Canary Islands and cite previous studies indicating polarization effects on the order of 1 part in $10^6$.
There is also literature on atmospheric polarization effects in the context of laser beam propagation for military purposes.
For example,  [\citenum{Korotkova04}] gives results for atmospheric polarization effects on a Gaussian Schell beam.
While it seems that the case of a plane wave incident upon the Earth from an astronomical source should be representable by a Gaussian Schell beam in the limit that the beam-width parameters go to infinity,  the laser propagation results given by  [\citenum{Korotkova04}] are unphysical in that limit. 


Given the current state of knowledge of atmospheric polarization effects, it will be assumed that the atmosphere contributes no polarization aberration, so the star's Stokes parameters are known and correspond to those that can be measured in a standard observational setting (with long exposures that average over the turbulence), which is likely to be correct to first order.
Thus, the effect of the atmospheric turbulence will be considered to be a scalar phenomenon described by a complex-valued phase screen (imaginary values account for scintillation) $\phi_\ra(\br,t)$.
Then, as there are no optics between the $-1$ plane (above the atmosphere) and the $0$ plane (the telescope entrance pupil), the Jones vector of the starlight at the entrance pupil is given by:
\begin{align}
u_{\star 0}(\br,t) & = u_{\star -1}(\br) \exp \big[ j \phi_\ra(\br,t) \big] \nonumber \\
& =  \sqrt{I_\star \,}
\exp j \big[ k  \balpha_\star \cdot \br  +  \phi_\ra(\br,t) \big] \breve{u}_\star \, ,
\label{u_star_0}
\end{align} 
which makes use of Eq.~(\ref{u_star_above_atmos}).

Some remarks on the limitations of Eq.~(\ref{u_star_0}) are in order.  
As mentioned in the discussion leading to Eq.~(\ref{E_complicated}), the local plane of polarization at position $\br = \bp$ must have a dependence on turbulent modulation.  
As a thought experiment, consider atmospheric fluctuations in the telescope entrance pupil plane in a small vicinity of the point $\bp$ at some time $t_0$\ resulting in tilt only, so that $\phi_\ra(\br,t_0)|_{\bp} = \alpha x$ [where $\br = (x,y)$, and $\tan^{-1} \alpha$\ is the tilt angle].
Since the tilt changes the direction of propagation, an unpolarized beam would have an electric field component parallel to the $z$-axis, while the $x$-component would be correspondingly reduced. 
Consider that typical seeing conditions in a world-class observatory correspond deviations in the angle of propagation on the order of 1 arc second ($''$).
A deviation $\alpha = 1''$\ would correspond to a $z$-component of the electric field amplitude of about $5 \times 10^{-6}$\ time the $x$-component, but impact on the $x$-component itself would only be about 1 part in $10^{11}$.   
While these numbers are not large, one must keep in mind that the telescope has optics that reduce the beam diameter while amplifying the angle.  
Consider an ELT that reduces the beam from a primary diameter of 50 m to a diameter of 2 cm (a factor of 2500) by the time it impinges on the DM, making $\alpha = 2500'' \approx 0.7^\circ$\ on the internal pupil plane corresponding to the DM.  
Thus, the atmospheric fluctuations combined with the high demagnification correspond to a local beam wobble on the order of $1^\circ$ in the vicinity of the conjugate point of $\bp$. 
This deviation now imparts $z$-component of over $1\%$\ of the original $x$-component of the electric field amplitude, which itself would be reduced by nearly 1 part in $10^4$.  
Such considerations may not be negligible in precision optical modeling and require further investigation.
    
Using Eq.~(\ref{u_star_0}), the star's mutual coherency at the entrance pupil is:
\begin{align}
\gamma_{\star 0}(\br_1, \br_2,t)   
  & =  \gamma_{\star -1}(\br_1,\br_2) \exp j \big[ \phi_\ra(\br_1,t) - \phi_\ra(\br_2,t) \big] \, ,
\label{gamma_star_0-1} \\
& =  J_\star  \exp j \big[ \phi_\ra(\br_1,t) - \phi_\ra(\br_2,t) + k \balpha_\star \cdot (\br_1 - \br_2)  \big] 
 \label{gamma_star_0}   
\end{align}
where Eq.~(\ref{gamma_star_0}) uses Eq.~(\ref{VcZ_star}).
From Eq.~(\ref{gamma_star_0}), the star's coherency at the entrance pupil is simply:
\begin{equation}
J_{\star 0}(\br,t) 
=   J_{\star } \exp \big[-2 \Im \big( \phi_\ra(\br,t) \big) \big]  \, .
\label{J_star_0}
\end{equation}

To calculate the mutual coherency of the planetary light, one can assume that its angular size is small enough that anisoplanitism effects are negligible, so that the planetary field is modulated by the same factor $\exp[ j \phi_\ra(\br,t) ]$.
Any small deviations from this assumption should be inconsequential, as they would result in a slight blurring of the planetary component of the image, not the stellar component.
As there is no convenient factorization of the mutual coherency of the planetary light in the $-1$ plane above the atmosphere, as in Eq.~(\ref{u_star_above_atmos}), on can multiply $u_{\rp-1}$\ in Eq.~(\ref{u_p-1}) by  $\exp[ j \phi_\ra(\br,t) ]$\ and follow the same steps to arrive at a blurred version of Eq.~(\ref{vCZ}) (which is simple because the atmospheric modulation factor does not depend on $\balpha$), to find:
\begin{equation}
\gamma_{\rp 0}(\br_1,\br_2,t) 
  =  \frac{1}{\pi} \exp j \big[ \phi_\ra(\br_1,t) - \phi_\ra(\br_2,t)  \big] 
   \int_\rp   \rd \balpha \,
S(\balpha)  \exp \left[ j k \balpha \cdot (  \br_1 - \br_2   ) \right] \, .
\label{vCZ_blurred}
\end{equation}
Eq.~(\ref{vCZ_blurred}) shows that the Fourier transform of the polarimetric planetary image, $S(\balpha)$, is multiplied by a complex-valued atmospheric modulation function, $ \exp j \big[ \phi_\ra(\br_1,t) - \phi_\ra(\br_2,t)  \big]$.
Thus, in an imaging system, which must have Fourier transforming properties, the $S(\balpha)$ will be convolved with the Fourier transform of  $ \exp j \big[ \phi_\ra(\br_1,t) - \phi_\ra(\br_2,t)  \big]$, resulting in an  image that is distorted by the atmospheric turbulence.

\begin{figure}[t]
\includegraphics[width=.45\linewidth,clip=]{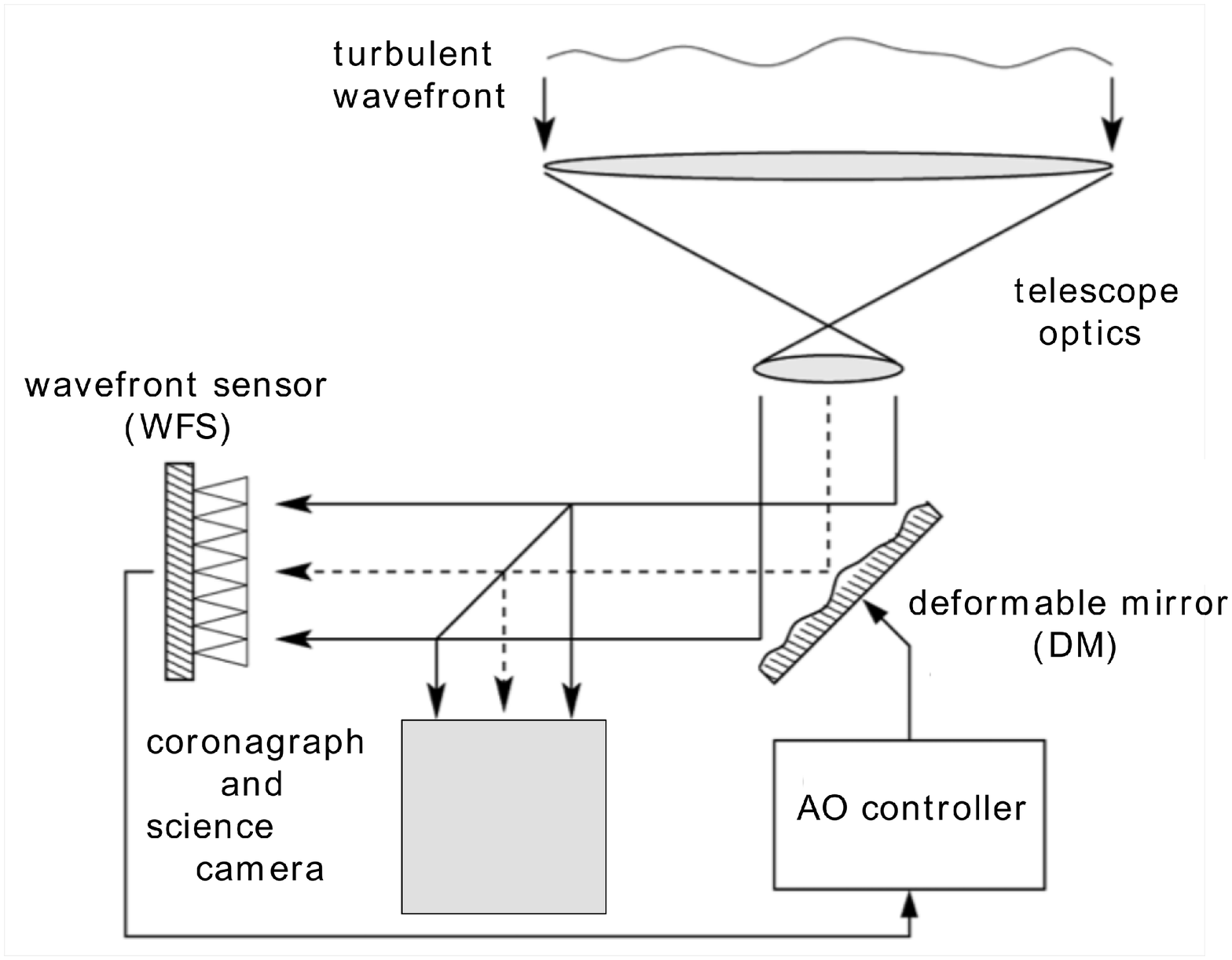}
\caption{\small Schematic diagram of an optical system containing a telescope with a closed-loop AO system and a coronagraph.  Modified from [\citenum{Hinnen_H2control}].}
\label{fig_schematic}
\vspace{-4mm}
\end{figure}

\subsection{Through the Optical System}\label{throughOS}

Consider a ground-based telescope equipped with a WFS and coronagraph, as shown schematically in Fig.~\ref{fig_schematic}.
The telescope, coronagraph and the science camera comprise a system consisting of a number of optical surfaces, with the telescope entrance pupil corresponding to surface number $0$ and the science camera detector surface having index $C$.
Let $\Upsilon_{l+1,1}(\br_{l+1},\br_l;\btheta_l)$, where $\br_l$\ and $\br_{l+1}$\ are the coordinate vectors in planes $l$ and $l+1$, respectively, be a Jones propagator from surface $l$ to surface $l+1$, including any aperture geometry or other structure on surface $l$.
The $\Upsilon_{l+1,1}$ kernel explicitly excludes interaction with the $l+1$\ surface.
The vector $\btheta_l$ is a set of experimentally determined parameters that modify $\Upsilon_{l+1,1}$ to account for various conditions such as alignment or other possibly significant factors.  

The field on the plane $l+1$ before interacting with the $l+1$ surface is given by integrating $u_l$ against the propagation kernel matrix: 
\begin{equation}
u_{l+1}(\br_{l+1},t) = \int \rd \br_l \, \Upsilon_{l+1,l}(\br_{l+1},\br_l;\btheta_l) u_l(\br_l,t)  \, .
\label{1-step}
\end{equation}
In order to be as general as possible and allow for unknown (scalar and polarization) aberration at each optical surface, the propagation kernel can be factored into known and unknown parts as follows:
\begin{equation}
\Upsilon_{l+1,l}(\br_{l+1},\br_l;\btheta_l) = 
\Upsilon_{l+1,l}^\mathrm{k}(\br_{l+1},\br_l;\btheta_l)   A_l^\mathrm{u}(\br_l) 
\label{Upsilon_general}
\end{equation}
where the "k" superscript stands for "known" and the "u" superscript stands for "unknown," $A_l^\mathrm{u}(\br_l)$\ is a $2 \times 2$\ Jones matrix function of unknown (possibly complex-valued) functions representing the unknown aberration caused by surface $l$.
If the $A_l^\mathrm{u}$\ function were parameterized, it could be included in the definition of the $\btheta_l$\ vector, as an alternative formulation.
Accounting for unknown aberrations in the form of Eq.~(\ref{Upsilon_general}) provides a bit of mathematical convenience as $A^\ru_l$\ can be regarded as a (small) deviation from the identity, i.e.:
\begin{equation}
A^\ru_l(\br_l) = \mathbb{I} + \tilde{A}^\ru_l(\br_l) \, ,
\label{A_tilde}
\end{equation}
where $\tilde{A}^\ru_l$\ is the deviation from the identity matrix ($\mathbb{I}$).
$A^\ru_l(\br_l)$\ is quite suitable for representing unknown scalar phase aberrations, such as small bumps on a mirror that have negligible effect on the polarization.  
Such a scalar aberration can be written as $A^\ru_l(\br_l) = \mathbb{I}\exp \big( j \phi_l^\ru(\br_l) \big)$, where $\phi^\ru_l$\ is the (possibly complex-valued) phase aberration on the optical surface $l$. 
Using the Taylor expansion of the exponential ($1 + j\phi_l^\ru(\br_l) - {\phi_l^\ru}^2 (\br_l)/2 \, +\cdots$), one has $\tilde{A}^\ru = \mathbb{I}\big( j\phi_l^\ru(\br_l) - {\phi_l^\ru}^2 (\br_l)/2 \, +\cdots  \big)$.
Then, using Eq.~(\ref{Upsilon_general}), the propagation kernel including an unknown a scalar aberration in plane $l$ can be written as:
\begin{equation}
\Upsilon_{l+1,l}(\br_{l+1},\br_l;\btheta_l) = 
\Upsilon_{l+1,l}^\mathrm{k}(\br_{l+1},\br_l;\btheta_l)  \mathbb{I} 
  [1 +  j\phi_l^\ru(\br_l) - {\phi_l^\ru}^2 (\br_l)/2 \, +\cdots ] \, .
\label{Upsilon_general_scalar}
\end{equation}

The choice of the kernel matrix $ \Upsilon_{l+1,l}$\ may be the Jones matrix function corresponding to an optical element in plane $l$ combined with some approximation of the Huygens-Fresnel principle to propagate the field to plane $l+1$.
For example, if the $2 \times 2$\ Jones matrix function for plane $l$ is given by $A_l(\br_l; \btheta_l)$ (which, if desired, can be factored into known and unknown parts as $A_l^\rk A_l^\ru$), the corresponding Huygens-Fresnel kernel can be written as:
\begin{equation}
\Upsilon_{l,l+1}(\br_{l+1},\br_l;\btheta_l) = \frac{-j k}{2 \pi} \, A_l(\br_l;\btheta_l) 
\exp [j k r(\br_{l+1},\br_l;\btheta_l) ] 
 \frac{\cos \vartheta(\br_{l+1},\br_l;\btheta_l)}{r(\br_{l+1},\br_l;\btheta_l)} \, ,
\label{HFkernel}
\end{equation}
where $r(\br_{l+1},\br_l;\btheta_l) $ is the Euclidean distance between the point $\br_l$ on surface $l$ and $\br_{l+1}$ on surface $l+1$, and $\vartheta(\br_{l+1},\br;\btheta_l)$\ is the angle between the aperture normal on surface $l$ and the line connecting these two points.

One can propagate the field from surface $l$ to surface $l+2$ by using the  kernel $\Upsilon_{l+2,l}$, which is defined as the contraction of the two kernels $\Upsilon_{l+1,l}$ and $\Upsilon_{l+2,l+1}$:
\begin{equation}
\Upsilon_{l+2,l}(\br_{l+2},\br_l;\btheta_{l+1},\btheta_l) \equiv 
\int \rd \br_{l+1}
\Upsilon_{l+2,l+1}(\br_{l+2},\br_{l+1};\btheta_l) 
 \Upsilon_{l+1,l}(\br_{l+1},\br_l;\btheta_{l+1}) \, .
\label{2-step}
\end{equation}
In operator notation Eq.~(\ref{2-step}) is written as:
\begin{equation}
\Upsilon_{l+2,l}(\br_{l+2},\br_l;\btheta_{l+1},\btheta_l) \equiv
\Upsilon_{l+2,l+1}(\br_{l+2},\br_{l+1};\btheta_l) 
 \Upsilon_{l+1,l}(\br_{l+1},\br_l;\btheta_{l+1}) \, ,
\label{2-step_operator}
\end{equation}
which implies integration over the coordinate that is the second argument in the first propagator and the first argument in the second propagator.

Similarly, the field from the star at the science camera, located in plane $C$\ can be determined from the propagation kernel $\Upsilon_{C,0}$, which is a contraction of the propagation kernels from plane $1$ through plane $C-1$:
\begin{equation}
\Upsilon(\br_C,\br_0; \bm(t) ;\btheta_{0:C-1}) \equiv 
{\displaystyle \int \rd \br_{C-1} \cdots \int \rd \br_1} \; 
 \Upsilon_{C,C-1}(\br_C,\br_{C-1}; \btheta_{C-1}) \cdots
 \Upsilon_{1,0}(\br_1,\br_0 \, ; \btheta_0) 
  \, ,
\label{1-to-C}
\end{equation}
where $\bm(t)$\ is a vector of DM commands, $\btheta_{0:C} \equiv [\btheta_0^\rT,\cdots,\btheta_C^\rT]^\rT$\ is a column vector of instrument parameters that collects the $\{\btheta_l \}$.
Note that in Eq.~(\ref{1-to-C}) hides the dependence on on $\bm(t)$\ insides the ellipsis ("$\cdots$"), which contains a propagation kernel corresponding to the DM that depends on $\bm(t)$. 
In operator notation, Eq.~(\ref{1-to-C}) is written as:
\begin{equation}
\Upsilon(\br_C,\br_0; \bm(t) ;\btheta_{0:C-1}) =  \prod_{l=0}^{C-1}
\Upsilon_{l+1,l}(\br_{l+1},\br_{l};\bm(t);\btheta_l) \, ,
\label{1-to-C_operator}
\end{equation}
in which integrations over the intermediate coordinate vectors are implicit.

Using Eq.~(\ref{u_star_0}), the field from the star at the science camera is given by integral:
\begin{equation}
u_C(\brho,t) = \sqrt{I_\star \,} \int \rd \br_0 \, \Upsilon_{C,0}(\brho,\br_0 \, ; \bm(t) ; \btheta_{0:C-1}) \breve{u}_\star 
  \exp j \big[ k \balpha_\star \cdot \br_0 + \phi_\ra(\br_0,t) \big] \, ,
\label{u_C_full}
\end{equation}
where the position in science camera plane is given by $\brho \equiv \br_C$.
The polarimetric image of the star can be determined with the help of Eqs.~(\ref{JonesPupilKron}) and (\ref{gamma_star_0}):
\begin{multline}
J_{\star C} (\brho,t) =  \int \rd \br_0 \int \rd \br_0'  \,
 \Upsilon_{C,0}(\brho,\br_0\, ; \bm(t) ; \btheta_{0:C}) \otimes \Upsilon_{C,0}^*(\brho,\br_0' \, ; \bm(t) ;\btheta_{0:C})  \times \\
  J_\star 
 \exp j \big[ \phi_\ra(\br_0,t) - \phi_\ra(\br_0',t) + k \balpha_\star \cdot (\br_0 - \br_0')  \big] \, .
 \label{star_image_full}
\end{multline}
Using Eq.~(\ref{vCZ_blurred}) , the polarimetric image of the planetary system (excluding the star) is given by:
\begin{multline}
J_{\rp C} (\brho,t) = 
\int \rd \br_0 \int \rd \br_0'  \exp j \big[ \phi_\ra(\br_0,t) - \phi_\ra(\br_0',t) \big] 
  \Upsilon_{C,0}(\brho,\br_0\, ; \bm(t); \btheta_{0:C}) 
\otimes \Upsilon_{C,0}^*(\brho,\br_0' \, ; \bm(t) ; \btheta_{0:C}) \; \times
\\   \int_\rp   \rd \balpha \,
S(\balpha)  \exp \left[  k \balpha \cdot (\br_0 - \br_0')  \right] \, .
\label{planet_image_full}
\end{multline}
Note that detailed propagation kernels in Eq.~(\ref{planet_image_full}) are likely to be unnecessarily precise, as much more simple approximations to calculate the planetary image should be adequate. 
This will be discussed in future papers in this series.
Adding Eqs.~(\ref{star_image_full}) and (\ref{planet_image_full}), the polarimetric image on the science camera detector is the sum of the stellar and planetary contributions:
\begin{equation}
J_C(\brho,t) = J_{\star C} (\brho,t) + J_{\rp C} (\brho,t) \, .
\label{J_C-total}
\end{equation}

Using the operator notation as in Eq.~(\ref{JonesPupilKronOperator}),  $\Upsilon_{C,0}(\brho,\br_0\, ; \bm(t); \btheta_{0:C}) 
\otimes \Upsilon_{C,0}^*(\brho,\br_0' \, ; \bm(t) ; \btheta_{0:C}) $\ also represents the operator that propagates the mutual coherency through the optical system.
Using Eq.~(\ref{1-to-C_operator}) and the identity (for suitably dimensioned matrices) $AC \otimes BD = (A\otimes B)(C \otimes D)$, this operator can be written as:
\begin{multline}
\Upsilon_{C,0}(\brho,\br_0\, ; \bm(t); \btheta_{0:C}) 
\otimes \Upsilon_{C,0}^*(\brho,\br_0' \, ; \bm(t) ; \btheta_{0:C}) \\
 = \: \prod_{l=0}^{C-1} 
\big[ \Upsilon_{l+1,l}(\br_{l+1},\br_{l}; \bm(t); \btheta_{l}) \otimes 
\Upsilon^*_{l+1,l}(\br'_{l+1},\br'_{l}; \bm(t); \btheta_{l})       \big] \, ,
\label{1-to-C-KronOperator}
\end{multline}
in which $\br'_C = \br_C = \brho$, and the $m(t)$\ (the DM command vector) argument only affects the operators in which the second coordinate corresponds to a DM.

Referring to Fig.~\ref{fig_schematic}, let the kernel that propagates the field from the telescope entrance pupil to the WFS entrance pupil be denoted as $\Upsilon_{w,0}(\br_w,\br_0 \, ; \btheta_{0:w-1} )$, where $w$\ is the index assigned to the plane corresponding to the WFS entrance pupil.
Neglecting the planetary contribution, the field in the WFS entrance pupil is given by:
\begin{equation}
u_w(\br_w,t) = \sqrt{I_\star \,} \int \rd \br_0 \, \Upsilon_{w,0}(\br_w,\br_0 \, ; \bm(t) ; \btheta_{0:w-1}) \breve{u}_\star 
  \exp j \big[ k \balpha_\star \cdot \br_0 + \phi_\ra(\br_0,t) \big] \, ,
\label{u_w_full}
\end{equation}
In Part 2 of this series, the SC field will be expressed in terms of $u_w$\ and non-common path aberrations, which are unique to the separate optical path experienced by the field impinging on the WFS. 

\section{Methods for Treatment of Aberrations}\label{Methods}

The purpose of this section is to outline the physical principles behind various aberration correction schemes in a rigorous fashion.
The level of rigor will allow the reader to understand the strengths and weaknesses of the various approaches.
This section does not discuss issues concerning noise, measurement errors, statistical or computational methodologies.   
These critical concerns are left to later papers in the series.

\subsection{Explicit Determination of Unknown Aberrations}\label{ExplicitDetermination}

The first step in solving for the unknown aberrations is to calculate how they manifest themselves in the polarimetric image. 
Using Eqs.~(\ref{Upsilon_general}) and (\ref{A_tilde}), and dropping terms that are 2\underline{nd} order or higher in $\{ \tilde{A}_l \}$, Eq.~(\ref{1-to-C_operator}) becomes:
\begin{multline}
\Upsilon_{C,0}(\br_C,\br_0; \bm(t) ) = \prod_{l=0}^{C-1}
\Upsilon^\rk_{l+1,l} \big( \br_{l+1},\br_{l}; \bm(t) \big) \; + \\
\sum_{k=1}^{C-1} \left( \prod_{l=k+1}^{C-1}  \Upsilon^\rk_{l+1,l} \big( \br_{l+1},\br_{l};\bm(t) \big)      \right) 
 \Upsilon^\rk_{k+1,k} \big (\br_{k+1},\br_{k};\bm(t) \big)  \tilde{A}^\ru_{k}(\br_k) 
 \left( \prod_{l=0}^{k-1}  \Upsilon^\rk_{l+1,l}\big( \br_{l+1},\br_{l};\bm(t) \big) \right) 
\label{AberOperator}
\end{multline}
where, for the sake of compactness, the $ \{ \theta_l \}$\ arguments, have been dropped, and in which the roman superscript $^\rk$, which stands for the word ``known,'' should not be confused with the index $k$\ (let alone the wavenumber $k = 2\pi/\lambda$).
In Eq.~(\ref{AberOperator}), the first term is simply the known part of the propagation operator, while the second term provides a 1\underline{st} order accounting of the surface aberrations $\{ \tilde{A}_k (\br_k) \}$ by summing over the effect of each surface.
In the second term, the rightmost factor propagates the field from the entrance pupil to the $k$\underline{th} surface, where the unknown aberration is located, the Jones matrix function $ \tilde{A}^\ru_{k}(\br_k)$\ applies the aberration, $\Upsilon_{k+1,k}$\ propagates the field to the $k+1$\ surface, and the leftmost factor propagates the field the rest of the way to the science camera detector surface.
Eq.~(\ref{AberOperator}) assumes that there are no unknown aberrations in at surfaces $0$ and $C$.
The first-order image plane manifestation of the unknown aberrations can be calculated from Eq.~(\ref{star_image_full}), using the expansion in Eq.~(\ref{AberOperator}), dropping terms that are 2\underline{nd} order in the $\{ \tilde{A}_k \}$\ functions.
In operator notation, the resulting polarimetric image of the star is:
\begin{multline}
J_{\star C}(\brho,t) \approx  \Bigg\{  \Upsilon^\rk_{C,0} \big(\brho,\br_0; \bm(t) \big) \otimes  
\Upsilon^{\rk *}_{C,0} \big(\brho,\br'_0; \bm(t) \big) + \\
 \Upsilon^\rk_{C,0} \big(\brho,\br_0; \bm(t) \big) \otimes
\Bigg[
\sum_{k=1}^{C-1} \left( \prod_{l=k+1}^{C-1}  \Upsilon^{\rk *}_{l+1,l} \big( \br'_{l+1},\br'_{l};\bm(t) \big)      \right) 
\Upsilon^{\rk *}_{k+1,k} \big (\br'_{k+1},\br'_{k};\bm(t) \big)  \tilde{A}^{\ru *}_{k}(\br'_k)
 \left( \prod_{l=0}^{k-1}  \Upsilon^{\rk *}_{l+1,l}\big( \br'_{l+1},\br'_{l};\bm(t) \big) \right) 
\Bigg] + \\
\Bigg[   \sum_{k=1}^{C-1} \left( \prod_{l=k+1}^{C-1}  \Upsilon^\rk_{l+1,l} \big( \br_{l+1},\br_{l};\bm(t) \big)      \right)
\Upsilon^\rk_{k+1,k} \big (\br_{k+1},\br_{k};\bm(t) \big)  \tilde{A}^\ru_{k}(\br_k) 
\left( \prod_{l=0}^{k-1}  \Upsilon^\rk_{l+1,l}\big( \br_{l+1},\br_{l};\bm(t) \big) \right)  \Bigg] 
\otimes \Upsilon^{\rk *}_{C,0} \big(\brho,\br'_0; \bm(t) \big) \Bigg\}  
\\ \times    J_\star  \exp j \big[ \phi_\ra(\br_0,t) - \phi_\ra(\br_0',t) + k \balpha_\star \cdot (\br_0 - \br_0')  \big]  \; ,
\label{StarImageAber}
\end{multline}
where $\Upsilon_{C,0}^{\rk} $\ is the first term in Eq.~(\ref{AberOperator}), corresponding to the known propagator, and $\br_C = \br'_C = \brho$.
Similarly to Eq.~(\ref{1-to-C-KronOperator}), the operators constituting $\Upsilon_{C,0}^\rk$\ may placed inside the serial products over the indicies $l$.  
Eq.~(\ref{StarImageAber}) is linear in unknown aberration functions  $\{ \tilde{A}_k (\br_k) \}$ and is the fundamental equation for statistical inference of their values, under the assumption that the aberrant field is small enough that the 2\underline{nd} order interactions are negligible.
This assumption can be improved once the largest unknown aberrations are determined by including them in the known propagation operators.
It is also important to note that Eq.~(\ref{StarImageAber}) makes no assumptions about the DM commands, which may include deliberate offsets associated with speckle nulling or electric field conjugation schemes, as discussed in Sec.~\ref{Methods}\ref{EFC}.

In Part 2 of this series, it will be shown that the SC image can be written in terms of $u_w$, the field measured by the WFS.  
The resulting expression eliminates the need to consider aberrations upstream of the beam splitter, which separates the optical paths of the SC and WFS, and the result is a versions of Eqs.~(\ref{AberOperator}) and (\ref{StarImageAber}) with many fewer unknown functions.

Of course, the image in the science camera also includes the planetary image, as per Eqs.~(\ref{planet_image_full}) and (\ref{J_C-total}).  The known propagation operator $ \Upsilon^\rk_{C,0} $\ should be more than adequate for calculating the planetary image.
Thus, when one combines Eq.~(\ref{StarImageAber}) with a version Eq.~(\ref{planet_image_full}) that utilizes  $ \Upsilon^\rk_{C,0} $, the resulting expression for the polarimetric image in the science camera is linear in both the unknown aberration functions and the planetary image $S(\balpha)$.
Then, the exoplanet imaging problem in the fairly general form developed here can be stated as:
Given data streams corresponding to measurements of the residual phase $\phi_\rr(\br,t)$\ in Eq.~(\ref{u_w_full}) and $J_C(\brho,t)$\ in Eq.~(\ref{J_C-total}), estimate $S(\balpha)$, $\btheta_{0:C-1}$, and $\{ \tilde{A}^\ru_l(\br_l) \}$.
As the image $J_C$\ is strongly modulated by the atmospheric fluctuations $\phi_\ra$, the use of the WFS data stream is critical to success in this endeavor.\cite{Frazin13,Frazin14}
While estimating so many unknown functions is certainly a daunting challenge, it should not be dismissed so readily, as large millisecond-cadence data sets, modern hardware, and computational$/$statistical methods (e.g., sparse representations) could make significant strides in this direction a realistic long-term goal.   
In the short term, simplifications must be considered, which will be treated in future papers in this series.

\subsubsection{The Concept of Equivalent Aberrations}\label{EquivAber}

Determining unknown aberration functions on many optical surfaces simultaneously is an ambitious goal, and finding methods to reduce the number of unknown functions that need to be determined is worth some effort.
Consider a hypothetical optical system with unknown aberrations in only two planes, $M$\ and $N$, where $N > M > D$, meaning that plane $N$ is downstream of plane $M$, which is downstream of the DM plane $D$.
Assume that the $M$ and $N$ planes have small scalar aberrations, so that, as per Eq.~(\ref{A_tilde}), $\tilde{A}^\ru_M = \mathbb{I} j \phi^\ru_M(\br_M)$\ and $\tilde{A}^\ru_N = \mathbb{I} j \phi^\ru_N(\br_N)$.  
Under what circumstances can one account for the aberration in plane $M$ by replacing it with some equivalent aberration in plane $N$?
Using Eq.~(\ref{Upsilon_general_scalar}), the aberrant part of the field incident on plane $N$\ is:
\begin{equation}
u^\ru_N(\br_N,t) =
j \int \rd \br_M \,  \Upsilon^\rk_{N,M} \big(\br_N,\br_M \big)  \phi_M^\ru(\br_M)  u_M(\br_M,t) \,
\label{u_m-to-u_n}
\end{equation}
assuming that there are no active optical elements (i.e., DMs) between $M$ and $N$.
Replacing $ \phi_M^\ru(\br_M) $\ with some functionally equivalent aberration in the $N$\ plane equates making the following assertion:  There exists some complex-valued and time-independent function $\psi(\br_N)$ such that, for all (reasonable) values of $u_M(\br_M,t)$ and $ \phi_M^\ru(\br_M) $, the following relation holds:
\begin{equation}
 \psi(\br_N)   \int \rd \br_M \,  \Upsilon^\rk_{N,M} \big(\br_N,\br_M \big)  
   u_M(\br_M,t) = 
 \int \rd \br_M \,  \Upsilon^\rk_{N,M} \big(\br_N,\br_M \big)  \phi_M^\ru(\br_M)
   u_M(\br_M,t) \, .
\label{some_psi}
\end{equation}
If this relation holds true, the $k=M$\ term in Eq.~(\ref{AberOperator}) can be omitted if the $k=N$\ term is modified according to $ \phi^\ru_N(\br_N) \rightarrow  [ \phi^\ru_N(\br_N) + \psi(\br_N) ]$.  
From Eq.~(\ref{some_psi}), $\psi(\br_N)$\ would be given by the ratio:
\begin{equation}
 \psi(\br_N) = \frac{\int \rd \br_M \,  \Upsilon^\rk_{N,M} \big(\br_N,\br_M \big)  \phi_M^\ru(\br_M)
   u_M(\br_M,t)}{\int \rd \br_M \,  \Upsilon^\rk_{N,M} \big(\br_N,\br_M \big)  u_M(\br_M,t)} \, ,
\label{some_psi_ratio}
\end{equation}
possible zeros in the denominator not withstanding.
In order for the time-independent function $\psi(\br_N)$\ to exist, one of several conditions must be met:
\begin{itemize}
\item{There are no unknown aberrations, so $\phi^\ru_M = 0$.}
\item{The incident field is not time-dependent so $u_M(\br_M,t) = u_M(\br_M)$.
If the $M$\ plane is after the DM, condition is approached as the quality of the AO correction improves, i.e., when a high Strehl ratio is achieved.}
\item{The propagation operator $\Upsilon^\rk_{N,M}$\ is behaves much like an identity operator, so that for some arbitrary field $u_M(\br_M)$, (in operator notation) $u_N(\br_N) =  \Upsilon^\rk_{N,M} \big(\br_N,\br_M \big) u_M(\br_M) = u_M(\beta \br_N)$\ for some (real-valued) magnification factor $\beta$, in which case $\psi(\br_N) = \phi^\ru_M(\beta \br_M)$.
Under geometrical optics approximations, this condition is achieved when the $M$\ and $N$\ planes are conjugate.}
\end{itemize}
There may be other, most likely contrived, functional forms of $\Upsilon^\rk_{N,M}$\ and $\phi^\ru$\ in which $\psi(\br_N)$ exists.
The general formula for propagating the field from one surface to the next involves the Huygen-Fresnel kernel, shown in Eq.~(\ref{HFkernel}), which does not behave like the identity operator unless intermediate optical surfaces create $2^n$ (where $n$ is a natural number) Fourier transforms, or the beam is columnated and propagation effects can be neglected.
Therefore, it seems that the treating aberrations on multiple optical surfaces equivalently to aberrations on a single optical surface, is an approximation that needs to be evaluated on a case-by-case basis.

\subsection{Black Box Models}\label{BlackBox}

The developments thus far allow the reader to understand the speckle cancellation method of 
 [\citenum{Martinache_SpeckleCancel12}] and the background estimation method of  [\citenum{Codona13}].
As per Eq.~(\ref{AberOperator}), the end-to-end propagator can be written as the sum of known and unknown propagators: $\Upsilon_{C,0} = \Upsilon^\rk_{C,0} + \Upsilon^\ru_{C,0}  \, $, where the unknown propagator is the second term in Eq.~(\ref{AberOperator}), plus terms that are nonlinear in the $\{ \tilde{A}_k \}$, if desired. 
Here, this is called the "black box model" since the unknown aberrations are only taken into account via the resulting image-plane field $u^\ru  $.
Black box models have the considerable advantage of not requiring a sophisticated model of the optical system and a parameterization of the unknown aberrations.
Assume that the WFS data stream leads to an estimate of the atmospheric modulation $\exp [ j \hat{\phi_\ra}(\br_0,t)]$.  
Then, the field in the science camera is given by:
\begin{equation}
u_C(\brho,t) \approx u^\rk(\brho,t) + u^\ru(\brho,t) \, ,
\label{CodonaField}
\end{equation}
in which the known and unknown parts of the field are given by [see Eq.~(\ref{u_C_full})]:
\begin{align}
u^\rk(\brho,t) = & \;  \sqrt{I_\star \,} \int \rd \br_0 \, \Upsilon_{C,0}^\rk \big( \br_C,\br_0; \bm(t) \big)
 \breve{u}_\star   \exp  \big[ j \hat{\phi_\ra}(\br_0,t) \big]
\label{CodonaKnown} \\
u^\ru(\brho,t) = & \;  \sqrt{I_\star \,} \int \rd \br_0 \,  \Upsilon_{C,0}^\ru \big( \br_C,\br_0; \bm(t) \big) 
 \breve{u}_\star   \exp  \big[ j \hat{\phi_\ra}(\br_0,t) \big] 
\label{CodonaUnknown} \,
\end{align}
in which $\balpha_\star = 0$\ has been assumed, and the $\btheta_{0:C}$\ argument has been dropped for simplicity. 
The method of  [\citenum{Codona13}] requires the unknown field $u^\ru$\ (``the subject beam'') to be independent of the atmospheric modulation (and therefore time), i.e., $ u^\ru(\brho) \approx  \sqrt{I_\star \,} \int \rd \br_0 \, \Upsilon_{C,0}^\ru(\br_C,\br_0) \breve{u}_\star $, which is not consistent with Eq.~(\ref{CodonaUnknown}).
Then, ignoring the planetary contribution and polarization effects, the image in the science camera is given by
\begin{eqnarray}
I_C(\brho,t) & (=) & u_C(\brho,t)  u^*_C(\brho,t)  \nonumber \\
 & (=) & u^\rk(\brho,t)  u^{\rk *}(\brho,t) + 2 \Re[u^\ru(\brho)  u^{\rk *}(\brho,t) ] 
 + u^\ru(\brho)  u^{\ru *}(\brho) \, ,
 \label{CodonaCoherency}
\end{eqnarray}
where $I_C$\ is the intensity measured in the science camera, $u^\ru$ and $u^\rk$\ are taken to be scalar fields, $\Re$\ indicates that the real part is to be taken, and the $=$\ is placed inside parentheses because the equation is based on the incorrect assumption discussed above.
Given a time-series of measurements of $J_C(\brho,t)$\ and estimates $\hat{\phi_\ra}(\br,t)$, Eq.~(\ref{CodonaCoherency}) allows determination $u^\ru(\brho)$ and, therefore, the intensity in the image plane arising due to unknown aberrations, i.e., $2 \Re[u^\ru(\brho)  u^{\rk *}(\brho,t) ] + u^\ru(\brho)  u^{\ru *}(\brho)$.
With $u^\rk(\brho,t)$\ and $u^\ru(\brho)$\ in hand, the background to be subtracted from the target image is simply the time-integral of $|| u^\rk(\brho,t) + u^\ru(\brho) ||^2$.

Of course, the major criticism of this approach is the assumption that the field arising from the unknown aberrations is independent of the  turbulent modulation, which, clearly, it is not, according to Eq.~(\ref{CodonaUnknown}). 
Indeed, the real part $\phi_\ra$\ may well vary over multiples of $2 \pi$\ over the pupil.  
However, this assumption is improved by the fact that, for planes in the optical system succeeding the DM, the uncorrected portion of $\phi_\ra$, the residual phase $\phi_\rr$, has a significantly smaller amplitude.
It is likely that the unknown aberrations are most significant after the DM, because, in a closed-loop system, the AO system corrects most of the aberrations up to the DM plane.

There may be a possibility to improve the method of  [\citenum{Codona13}] for closed-loop AO systems by assuming that the unknown portion of the field in the image plane is a function only of the residual phase $\phi_\rr(\br,t)$, so $u_\ru(\brho) \rightarrow u^\ru\big( \brho;\phi_{\br}(\br,t) \big)$.
Then for small amplitudes of $\phi_{\br}(\br,t)$, one has
\begin{equation}
 u^\ru\big( \brho;\phi_{\br}(\br,t) \big) \approx u^\ru\big( \brho; 0  \big)
 + \frac{\partial u^\ru\big( \brho;\phi_{\br}(\br,t) \big)}{\partial \phi_{\br}} \biggr|_{\phi_\rr = 0} 
 \hat{\phi_\rr}(\br,t) \, ,
 \label{CodonaDeriv}
\end{equation}
which makes use of the functional derivative and WFS estimate of the residual phase $\hat{\phi_\rr}$.
Then one could replace $u^\ru(\brho)$\ in Eq.~(\ref{CodonaCoherency}) with $u^\ru\big( \brho;\phi_{\br}(\br,t) \big)$\ in Eq.~(\ref{CodonaDeriv}), and use measurements to solve for both $u^\ru\big( \brho; 0  \big)$\ and a representation of the functional derivative $\frac{\partial u^\ru\big( \brho;\phi_{\br}(\br,t) \big)}{\partial \phi_{\br}} \bigr|_{\phi_\rr = 0} \, $.
The difficulty of this scheme increases dramatically if higher order functional derivatives are needed to approve the approximation in Eq.~(\ref{CodonaDeriv}).

The speckle cancellation method of  [\citenum{Martinache_SpeckleCancel12}] does not make use of the WFS data stream or millisecond exposures in the science camera.
The goal of the speckle cancellation procedure is to use the DM to cause starlight to destructively interfere with an undesired speckle at position $\brho$. 
This procedure is repeated to remove multiple speckles at various locations.     
In this method one seeks to cancel out a speckle in the image plane at the location $\brho$, caused by an aberration leading to speckle field $ u^\ru ( \brho )$, making the same assumption as  [\citenum{Codona13}], i.e., ignoring its dependence on the atmospheric turbulence, in violation of Eq.~(\ref{CodonaUnknown}).
In this method, relatively long exposures are used, so the observations are effectively time-averages over the atmospheric turbulence.
The principle of the method is to use a several known DM offsets ${\bm_k}$, in which the index $k = (0,1,3)$\ corresponds to the phases $\{ \varphi_k \} = (0,\pi/2,\pi,3\pi/2)$ of a sinusoidal DM pattern corresponding to the spatial frequency $\brho$.  
Then, at a given time $t$\ the DM position corresponds to $\bm(t) + \bm_k$, in which $\bm(t)$\ is the set of commands determined by the AO control loop. 
Then the known part of the field is $u^\rk(\brho,t;\varphi_k)$, which can be calculated from Eq.~(\ref{CodonaKnown}) using $\bm(t) + \bm_k$\ as an argument inside $\Upsilon^\rk_{C,0}$ instead of $\bm(t)$.
As long exposures are used, one measures $\overline{I_C}(\brho,t;\varphi_k)$, where the overline indicates temporal averaging over the atmospheric turbulence, to determine the amplitude and phase of $u^\ru(\brho)$.  
Once the amplitude and phase of $u^\ru(\brho)$\ have been determined, the DM is commanded to implement a canceling sinusoidal aberration $\delta \bm$ that gives rise to a field $u^\rc(\brho)$\ that is equal in amplitude, but opposite in phase to, $u^\ru(\brho)$.  
In reality neither $u^\ru$\ nor $u^\rc$\ will be independent of the atmospheric modulation, but some cancellation can be achieved, and  [\citenum{Martinache_SpeckleCancel14}] demonstrates considerable success on-sky with this technique.

\subsection{Dark Hole Generation \& Electric Field Conjugation}\label{EFC}

Considerable effort has been devoted to using the DM (or multiple DMs) to minimize (or otherwise control) the starlight in some specified region of the image plane.
The history of these efforts is reviewed in  [\citenum{Pueyo_EFC09}] and  [\citenum{Martinache_SpeckleCancel12}].
The idea behind so-called "dark hole" methods is to minimize the intensity integrated over a control region of interest,\cite{Traub_Nulling06} which requires first determining the speckle field using prescribed DM offsets, much as described above in Sec.~\ref{Methods}\ref{BlackBox} in connection with the speckle cancellation method of  [\citenum{Martinache_SpeckleCancel12}].
So-called "electric field conjugation" does not seek to minimize the intensity in the control region.
Rather, its objective is to make the field what it would be in the absence of unknown aberration, or, in other words, to attain some ideal diffraction pattern.   
Modern implementations set the DM offset command vector $\delta \bm$\ to a value that minimizes the required DM stroke while the desired image plane electric field is used as a constraint on the optimization.\cite{Pueyo_EFC09,Kasdin_EFC13}

Dark hole and field conjugation methods assume that the unwanted speckle in the image plane is caused by a complex-valued unknown aberration in some plane of the optical system, most often a pupil plane corresponding to the location of a DM.
Let us denote this plane with the index $N$.
If the unknown aberrations in the optical system are truly confined only to the $N$\ plane, then the field in the science camera is determined by the propagation kernel in Eq.~(\ref{AberOperator}), keeping only the $k=N$\ term in the summation.
Thus, one fundamental limitation with these methods is that they ignore the $k \neq N$ terms, and implicitly assume that all of the unaccounted aberration can be treated by an equivalent pupil plane aberration, as discussed in Sec.~\ref{Methods}\ref{EquivAber}.
In observations from space, the assumption of equivalent aberrations may be justified, but the ratio in Eq.~(\ref{some_psi_ratio}) will have some (relatively slow) time-dependence due to telescope dynamics, requiring the DM solution to the field conjugation problem to be updated regularly.\cite{Kasdin_EFC13}
If these updates are accurate and on a time-scale that is short enough, the approximation of equivalent aberrations should not prove to be a problem.
In ground-based astronomy, this approximation of equivalent aberrations needs to be evaluated with much greater care and will depend not only on propagators between the planes but also on the assumed quality of the AO correction.

One potential improvement would be to generalize dark hole and field conjugation methods to treat aberrations in multiple planes using multiple DMs, but this has not been explored, to the best of this author's knowledge. 
Using a complicated series of DM offsets to determine the speckle fields arising from aberrations in multiple planes may well introduce non-trivial complications, but perhaps such an approach could be unified with the methods described in this series of papers.

\section{Conclusions}

Direct imaging of exo-planetary systems from ground-based telescopes is fraught with technical challenges, but continuing advances in optical technology such as ultra-low noise, rapid-readout optical and infrared detectors are opening new avenues of opportunity.
Taking the greatest advantage of these technologies will require harnessing simultaneous wavefront sensor and science camera data streams at millisecond cadences to differentiate the planetary signal from the aberrant starlight.
The work presented here is Part I of a statistical framework that should prove useful for self-consistently determining the unknown aberrations and their effects on the star's image in a polarizing optical system, while simultaneously solving for the (polarimetric) planetary image. 
The formalism presented here provides a new perspective on other methods that utilize focal plane wavefront sensing, and shows that their simplifying assumptions may limit their effectiveness in ground-based astronomy.
In particular, current dark hole and electric field conjugation techniques assume that the speckle field arises due to aberrations in some specific (pupil) plane of the optical system, tacitly presuming that all of the aberrations on the various optical surfaces can be represented as a complex-valued aberration in that plane.  
Sec.~\ref{Methods}\ref{EquivAber} shows that this assumption of equivalent aberrations cannot be true in a strict mathematical sense, limiting the effectiveness of these methods, but the accuracy and consequences of this assumption needs to be evaluated on case-by-case basis.
In principle, reliance on the assumption of equivalent aberrations could be lessened by generalizing field conjugation and dark hole methods to correct for aberrations in multiple planes.
Future techniques that use DMs to correct for multi-plane aberrations would likely benefit from the developments in this series of papers for utilizing millisecond telemetry to determine the aberrations in multiple planes.
Despite the limitations of the current methods that utilize focal plane sensing, they be worth implementing to provide some correction of the speckle field, but tapping the ultimate potential of the optical system will likely require the more advanced techniques outlined herein and in future papers in this series.

An exoplanet imaging program that implements simultaneous determination of the image of the planetary system and aberrations in multiple planes, utilizing millisecond telemetry from one or more WFSs and the science camera (and possibly including active aberration correction), is the most complete solution to the ultra-high contrast imaging problem.
Importantly, this solution paradigm can easily include constraints, such as those arising from the diurnal rotation (used by ADI), multi-wavelength imaging (using by SDI), and polarization of the light scattered by the planetary material (used in polarization differential imaging), leaving little reason to believe it cannot lead to a substantial improvement over current methods.
The chances for success of this new paradigm are greatly increased by the new generation of ultra-low noise, rapid-readout detectors.
However, practical implementation of such a program will require progress on many fronts, such as sophisticated optical modeling based on various approximations, evaluating the accuracy of those approximations, modeling the error statistics of wavefronts measured by WFSs, and development of specialized statistical signal processing algorithms.
These issues will be the subject of future papers in this series.

\section*{Acknowledgments} 
The author thanks Olivier Guyon, Wes Traub and Jim Breckinridge for encouragement and enlightening discussions.


\section*{Appendix: Formalism and Conventions}

Henceforth, this paper will adhere to the following conventions in order to minimize notation:
\begin{itemize}
\item{All time-variation made explicit in the notation corresponds to timescales of $\tau_\mathrm{G}$\ or longer.
Variations on timescales similar to $\tau_\mathrm{c}$\ or smaller are implicit and dropped from the notation. }
\item{When a function is independent of the spatial coordinate $\br$\ or the time $t$\ (corresponding to timescales of $\tau_\rG$\ or  longer), the argument is dropped.  The number of arguments needed will be clear from the context.}
\item{Averaging over many periods of $\tau_\rc$, as was done in Eqs.~(\ref{coherency_def}) and (\ref{mutual_coherency_def}) will be implicit whenever Jones vectors $u$\ are multiplied together.  Therefore, the $\langle \; \rangle_{\tau_\rc}$\ notation is dropped. }
\end{itemize}
With these conventions, the field due to an astronomical source above the atmosphere in Eq.~(\ref{polariz_factor2}) is written as $u(\br) = \sqrt{I \,} \exp [ j \phi(\br) ] \breve{u}(\br)$, but its coherency is independent of position and is written as $J = u(\br) \otimes u(\br)^* = I \breve{u}(\br) \otimes \breve{u}^*(\br)$.
Its mutual coherency is written as $\gamma(\br_1,\br_2) = u(\br_1) \otimes u^*(\br_2) = I \exp j [\phi(\br_1) - \phi(\br_2) ] \breve{u}(\br_1) \otimes \breve{u}^*(\br_2)$.
Now, assume the atmosphere applies a modulation to this field in the form of a (complex-valued) phase factor $\phi_\ra(\br,t)$, which varies on the $\tau_\mathrm{G}$\ timescale (see Sec.~\ref{PropModels}\ref{atmosphere}).
Then, the field observed on the ground is given by $u(\br,t) = \sqrt{I \,} \exp j [  \phi(\br) + \phi_\ra(\br,t)] \breve{u}(\br)$, and the corresponding coherency is $J(\br,t) = I \exp[ - 2 \Im \big( \phi_\ra(\br,t)  \big)  ] \breve{u}(\br) \otimes \breve{u}^*(\br) \,$.
Similarly, the mutual coherency on the ground is expressed as $\gamma(\br_1,\br_2,t)  = I \exp j [\phi_(\br_1) - \phi(\br_2) + \phi_\ra(\br_1,t) - \phi_\ra^*(\br_2,t) ]\breve{u}(\br_1) \otimes \breve{u}^*(\br_2)$.

In some of the developments in this series explicit integral notation will prove cumbersome, so operator notation will be preferred.   
Overloading the $\Upsilon$\ symbol so that it serves as both the propagation kernel as in Eq.~(\ref{JonesPupil}) and the operator that does the propagation should not introduce confusion, as the kernel needs to be accompanied by an integral sign to propagate the field, while in the operator notation the integration is implicit.\footnote{"Overloading" is a term borrowed from computer science.  An overloaded symbol has two or more meanings depending on the context.  For example, the "$+$" symbol is often overloaded, as it can be used to add numbers, matrices, functions, etc.} 
In operator notation, Eq.~(\ref{JonesPupil}) is written as:
\begin{equation}
u_1(\br_1,t) = \Upsilon(\br_1,\br_0) u_{0}(\br_0,t) \; ,
\label{JonesPupilOperator}
\end{equation}
and Eq.~(\ref{JonesPupilKron}) becomes
\begin{equation}
J_1(\br_1,t)  =    
 \Upsilon(\br_1,\br_0) \otimes \Upsilon^*(\br_1,\br_0')  \gamma_{0}(\br_0,\br_0',t) \, .
\label{JonesPupilKronOperator}
\end{equation}

\bibliography{exop}   
\bibliographystyle{osajnl}   

\end{document}